\documentclass[fleqn,10pt]{wlscirep}
\usepackage[utf8]{inputenc}
\usepackage[T1]{fontenc}
\usepackage{url}
\usepackage{booktabs}
\usepackage{multirow}
\usepackage{dsfont}
\usepackage{cleveref}
\usepackage{amsmath, bm}

\usepackage{color,soul}
\definecolor{lightblue}{rgb}{.90,.95,1}
\sethlcolor{white}
\usepackage{chngcntr}
\usepackage[symbol]{footmisc}

\title{Inequality and Inequity in Network-based Ranking and Recommendation Algorithms\footnote[2]{Please cite the \textbf{Scientific Reports} version of this paper: \url{https://doi.org/10.1038/s41598-022-05434-1}}}

\author[1,2,3]{Lisette Esp\'{i}n-Noboa} %
\author[4,5,1]{Claudia Wagner}
\author[6,4,1]{Markus Strohmaier} %
\author[1,*]{Fariba Karimi} %

\affil[1]{Complexity Science Hub, Vienna, Austria}
\affil[2]{Central European University, Vienna, Austria}
\affil[3]{University of Koblenz-Landau, Koblenz, Germany}
\affil[4]{GESIS -- Leibniz Institute for the Social Sciences, Cologne, Germany}
\affil[5]{RWTH Aachen University, Aachen, Germany}
\affil[6]{University of Mannheim, Mannheim, Germany}

\affil[*]{karimi@csh.ac.at}


\vspace{-10pt}
\begin{abstract}
Though algorithms promise many benefits including efficiency, objectivity and accuracy, they may also introduce or amplify biases. 
Here we study two well-known algorithms, namely PageRank and Who-to-Follow (WTF), and show to what extent their ranks produce %
\textit{inequality} and \textit{inequity} when applied to directed social networks.
To this end, we propose a \textbf{d}irected network model with \textbf{p}referential \textbf{a}ttachment and \textbf{h}omophily ({DPAH}) and demonstrate the influence of network structure on the rank distributions of these algorithms. Our main findings suggest that %
(i) inequality is positively correlated with inequity, 
(ii) inequality is driven by the interplay between preferential attachment, homophily, node activity and edge density, and 
(iii) inequity is driven by the interplay between homophily and minority size. 
\hl{In particular, these two algorithms \textit{reduce}, \textit{replicate} and \textit{amplify} the representation of minorities in top ranks when majorities are homophilic, neutral and heterophilic, respectively.
Moreover, when this representation is reduced, minorities may improve their visibility in the rank by connecting strategically in the network.
For instance, by increasing their out-degree or homophily when majorities are also homophilic.
These findings shed light on the social and algorithmic mechanisms that hinder equality and equity in network-based ranking and recommendation algorithms.} 
\end{abstract}

\begin{document}
\flushbottom
\maketitle
\thispagestyle{empty}

\vspace{-25pt}
\section*{Introduction}
Online social networks and information networks have become integral parts of our everyday life. 
However, the opportunities offered by such networks are often constrained not only by our previous interactions \cite{burt1976positions, coleman1988social, burt2003social, morselli2003career, bottero2011worlds}, but also by algorithms.
For instance, algorithms could make some people or content more visible than others via classification \cite{espin2021explaining}, ranking or recommendations \cite{abdollahpouri2019unfairness}.
In this regard, search engines and recommender systems are increasingly used for various applications such as whom to follow, whom to cite, or whom to hire.
Typically, these applications use algorithms to order items (e.g., people and academic papers) based on ``importance'' or ``relevance'', and may therefore produce social inequalities by discriminating certain individuals or groups of people in top ranks.
In fact, it has been shown that recommender systems such as \textit{Who-to-Follow} (WTF) \cite{gupta2013wtf} tend to increase the popularity of users who are already popular \cite{su2016effect,bellogin2017statistical,abdollahpouri2019unfairness}. A similar effect has been found in \textit{PageRank} \cite{page1999pagerank}, where nodes in high ranks stabilize their position and give little opportunity to other nodes to occupy higher positions \cite{ghoshal2011ranking}. 
This tendency towards the ``popular'' arises because these algorithms harness structural information, in particular, the in- and out-degree of nodes.
For this reason, modeling the directionality of links---which is often left out for simplicity---is crucial to really understand how these algorithms work on different types of networks. 

However, social networks are complex systems, and many other structural properties may also alter the distribution of nodes and groups in the ranking. 
For example, previous studies have shown that \textit{homophily}---the tendency to connect to similar others---affects the visibility of minorities in degree rankings \cite{karimi2018homophily} and people recommender systems \cite{fabbri2020effect}.
Consequently, it can reinforce societal issues such as the glass ceiling effect \cite{cotter2001glass, avin2015homophily, stoica2018algorithmic} and the invisibility syndrome \cite{franklin2000invisibility}. 
Despite these findings, little is known about the extent to which the combination of multiple structural properties can alter the visibility of minorities in top ranks from ranking and recommendation algorithms.
A further complication is that debiasing ranking outcomes and making them fair is very challenging
since they can be mitigated in different ways \cite{zehlike2021fairness}: by intervening on the score distribution of candidates \cite{kleinberg2018selection}, on the ranking algorithm \cite{asudeh2019designing}, or on the ranked outcome \cite{yang2017measuring}. 
While most of these studies tackle fairness in ranking, they do not explore the effects of networked data in ranking. This paper is a step towards this goal. Since such algorithms are so deeply involved in social, economic, and political processes, %
we need to first understand how our connections affect them to then apply appropriate interventions towards fair results.

\begin{figure}[ht!]
    \centering
    \includegraphics[width=\linewidth]{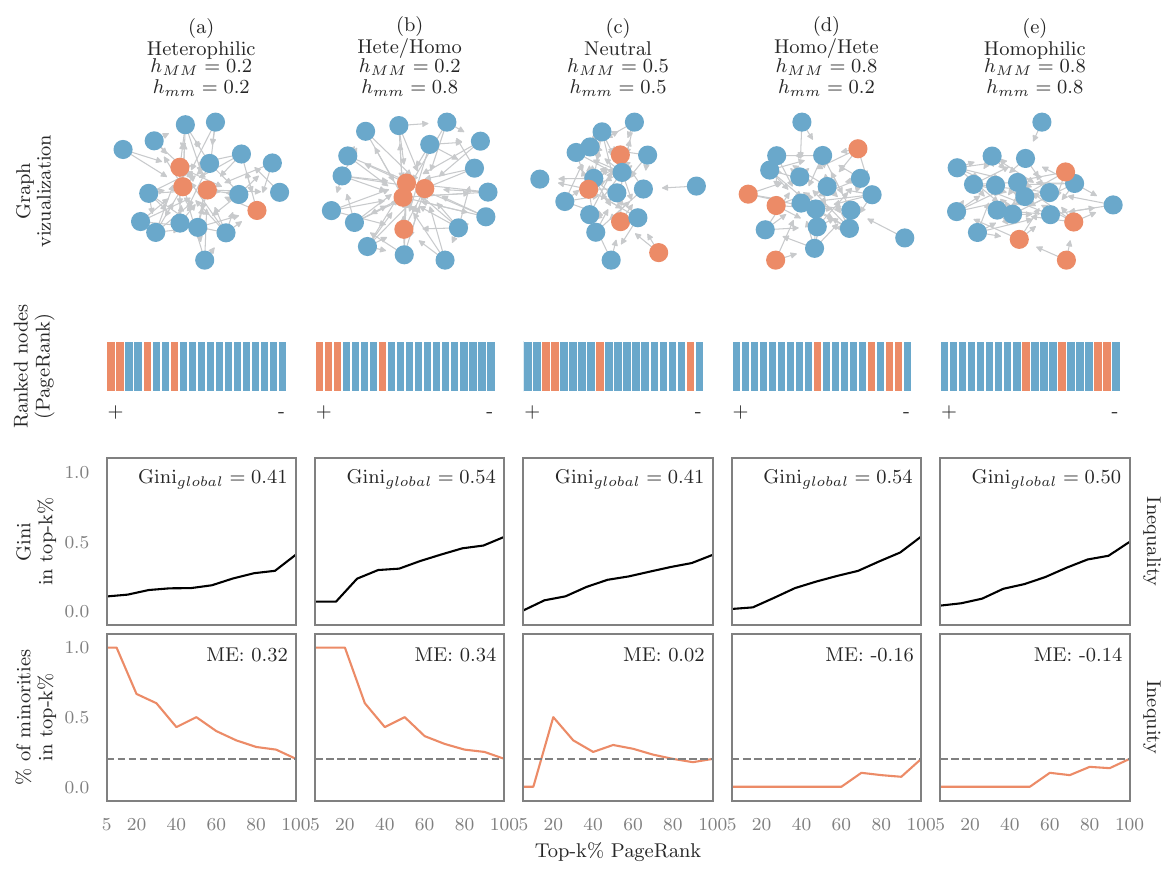}
    \caption{\textbf{Inequality and inequity.} Every column represents a network with certain level of homophily. All networks contain $20$ nodes: 20\% belong to the minority group (orange), and 80\% to the majority group (blue). Edges follow a preferential attachment with homophily mechanism. 
    The top row shows the graph and the level of homophily within groups ($MM$: majorities and $mm$: minorities). 
    The second row shows all nodes in descending order (from + to -) based on their PageRank scores. 
    The third row represents the rank \textit{inequality}: Gini coefficients of the rank distribution for every top$-k\%$ (black line). Gini$_{global}$ refers to the Gini coefficient of the entire rank distribution (i.e., at top-$100\%$).
    We see that the lower the $k$, the lower the Gini of the rank distribution.
    The bottom row represents the rank \textit{inequity}: Percentage of minorities found in each top-$k\%$ of the rank distribution (orange line). $ME$ is the mean error of these percentages compared to a fair baseline or diversity constraint (i.e., how much the orange line deviates from the dotted line across all top-k's). 
    Here we see three main patterns: 
    (a,b) When the majority group is heterophilic, minorities are on average over-represented, $ME>0.0$. 
    (d,e) When majorities are homophilic, minorities are on average under-represented, $ME<0.0$. 
    (c) When both groups are neutral, the observed fraction of minorities is almost as expected, $ME\approx 0$. }
    \label{fig:example}
\end{figure}

To this end, we propose {DPAH}, a network model that generates directed scale-free networks with binary-attributed nodes.
It encodes two main mechanisms of edge formation found in social networks: \textit{homophily} and \textit{preferential attachment}~\cite{borgatti2018analyzing, mcpherson2001birds, barabasi1999emergence} (see Methods for more details).
Moreover, it allows to control for the \textit{fraction of minorities}, \textit{edge density}, and the \textit{skewness of the out-degree distribution}.
By using this model, we systematically study how these structural properties of social networks impact the ranking of nodes in PageRank and WTF. 
In particular, we investigate two ranking
issues, inequality and inequity, and show how they get affected by the ranking algorithm together with the type of network.
We measure \textit{inequality} by quantifying the skewness of the rank distribution of nodes that PageRank and WTF produce, and \textit{inequity} as how well-represented the minorities are in the top of the rank compared to the proportion of minorities in the network.
In this work we study both %
ranking issues and measure their correlation.
Furthermore, we quantify them globally using the whole rank distribution, and locally within each top-k\% rank. The goal is to identify both the overall inequality and inequity trend that these algorithms produce, and the tipping points where minorities start gaining visibility in the top of the rank.

\begin{figure}[t!]
    \centering
    \includegraphics[width=0.7\linewidth]{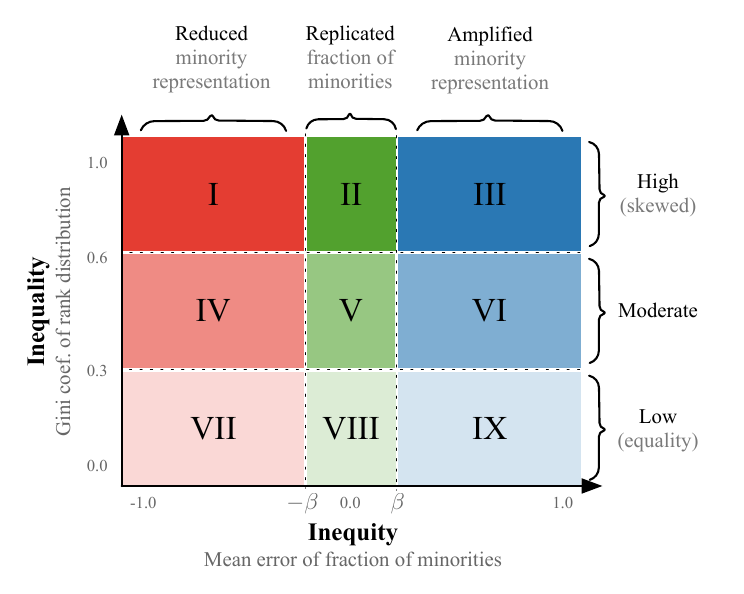}
    \caption{\textbf{Regions of disparity.} We measure \textit{inequality} (y-axis) as the skewness of the rank distribution, and \textit{inequity} (x-axis) as the mean differences between the proportional representation of groups in top-k\% ranks and the network.
    Highly skewed distributions lie in regions I to III (darker colors), and fair rankings, where minorities are well represented in top ranks, lie in regions II, V, VIII (green). 
    We set $\beta=0.05$ which is arbitrary and allows for a flexible region of \textit{group fairness}.}
    \label{fig:interpretation}
\end{figure}

As an example, consider the \textit{directed networks} shown in \Cref{fig:example}. Every column represents a network with two types of nodes, minority (orange) and majority (blue), and different levels of homophily within groups. Homophily $h$, is a parameter ranging from 0 to 1 and determines the tendency of two nodes of the same color to be connected. $h_{MM}$ and $h_{mm}$ represent homophily within majorities and minorities, respectively. When nodes are ranked using PageRank (second row), the position of the minorities in the rank varies \textit{systematically}. For instance, when majorities are heterophilic ($h_{MM}=0.2$, columns a and b), minorities often appear at the top (+). In contrast, when majorities are homophilic ($h_{MM}=0.8$, columns d and e), minorities tend to appear at the tail of the rank (-). Next, we explain this systematic ranking behavior in top ranks by further varying the structure of the network.

\section*{Results}

\subsection*{Inequality and inequity in ranking} %

\textit{Inequality} refers to the dispersion or distribution of \textit{importance} among \textit{individuals}. This importance is the ranking score assigned to every node by the algorithm.
We compute the \textit{Gini} coefficient of the rank distribution to measure how far the ranking scores of individuals deviate from a totally equal distribution (see Methods for more details).
As shown in \Cref{fig:interpretation}, a very low Gini score ($\text{Gini}<0.3$) means that %
individuals %
are very similar with respect to their ranking scores. If the Gini score is extremely high ($\text{Gini}\geq 0.6$), it means that only a few individuals capture most of the rank. In other words, the rank distribution is very skewed. Values in between ($0.3\leq\text{Gini}<0.6$) represent moderate skewed distributions.
Note that we measure inequality globally by using the whole rank distribution, and locally for each top-k\%.
From our example in \Cref{fig:example}, we see that PageRank on average generates moderate skewed ranking distributions for all the depicted networks ($Gini_{global}\approx0.5$). However, for very small top-k\%'s, the Gini is very low. This means that the top individuals possess very similar ranking scores.

\textit{Inequity} refers to \textit{group} fairness.
In particular, it measures the error distance between the fraction of minorities in the top-k\% and a given fair baseline (e.g., a diversity constraint or quota).
This baseline may be adjusted depending on the context of the application \cite{drosou2017diversity,singh2018fairness,zehlike2021fairness}. 
\hl{Here, a ranking is fair when its top-k\% preserves the proportional representation of groups in the network (i.e., equivalent to demographic parity}~\cite{dwork2012fairness, zehlike2021fairness}). Therefore, the error represents the local inequity at each top-k\%, and $ME$ the mean of these errors across all top-k\% ranks or global inequity.
As shown on the last row of \Cref{fig:example}, we measure the \textit{local inequity} in two steps. First, we compute the fraction of minorities that appear in each top-k\% rank (orange line). Second, we compute the error between the observed fraction of minorities in each top-k\% rank and a fair baseline (e.g., the actual fraction of minorities in the network, in this example $20\%$). Then, we average these error scores across all top-k\% ranks to determine the \textit{global inequity} score ($ME$ values). %
Ideally, a fair ranking should reach $ME=0$. However, in order to allow for small fluctuations we introduce the smoothing factor $\beta$. Thus, a fair ranking is such that $-\beta \leq ME \leq \beta$. %
The value of $\beta$ is arbitrary, and allows for a smooth definition of ``low mean error'' or fairness. We set $\beta=0.05$. 
As shown in~\Cref{fig:interpretation}, when $ME>\beta$, then minorities are over-represented in the top-k\% (blue region). 
When $ME<-\beta$, then minorities are under-represented (red region), otherwise %
the ranking is representing very well the minorities in the top of the rank (green region). %
Alternatively, we can say that the top rank (i) \textit{replicates} the proportional representation of groups when $ME$ is zero or very low, (ii) \textit{amplifies} the representation of minority nodes when $ME>\beta$, and (iii) \textit{reduces} the representation of minority nodes---and benefits the majority group---when $ME<-\beta$.
Note that $ME\approx 0$ may be an artifact of a numerical cancellation as in (c), the neutral case in \Cref{fig:example}. In such cases, we could argue that the ranking is still fair since overall it was biased towards both groups across all top-k's.

Finally, we refer to the relationship between inequality and inequity as \textit{disparity}. For example, if a ranking distribution achieves $Gini=0.65$ and $ME=0.5$, we say that the disparity lies in the region $III$ (dark blue), i.e., high inequality and high inequity, see \Cref{fig:interpretation}.

\begin{figure}[t!]
    \centering
    \includegraphics[width=1.0\textwidth]{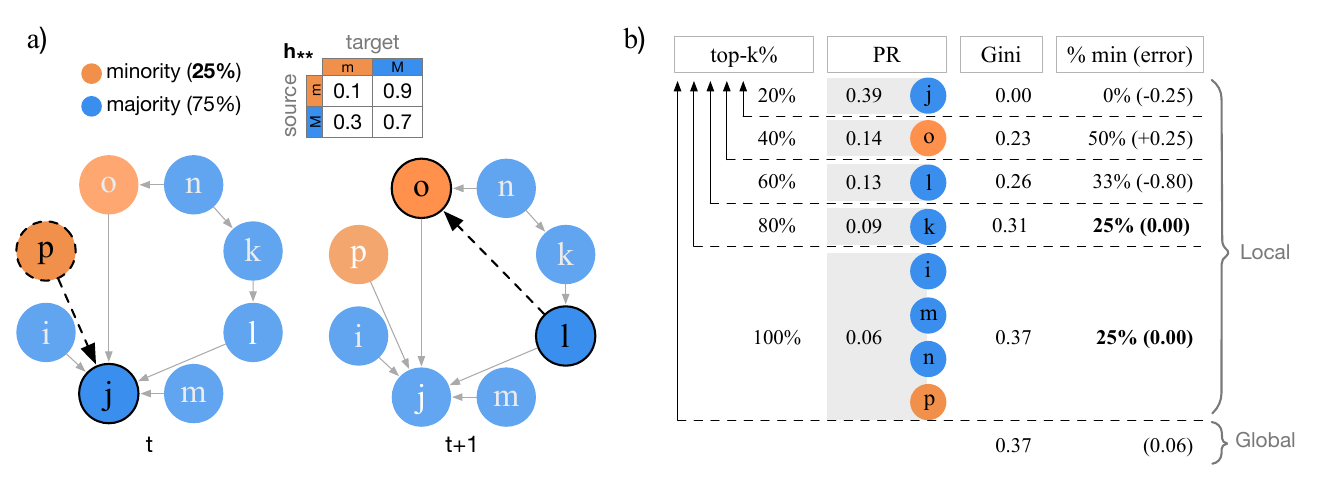}
    \caption{\textbf{{DPAH}~ model and ranking of nodes.} 
    a) Illustration of the directed network model with preferential attachment and homophily ({DPAH}). First, $n=8$ nodes are created and randomly labeled according to the fraction of minorities $f_m=0.25$. Then, the following algorithm repeats until a desired edge density is fulfilled. At time $t$, a source node $p$ is drawn from a power-law (activity) distribution, %
    and a target node $j$ is drawn with a probability proportional to the product of its in-degree $k_j^{in}=4$ %
    and the pair-wise homophily $h_{pj}=h_{mM}=0.9$. At time $t+1$, a new edge is added between %
    nodes $l\rightarrow o$ based on the same mechanism.
    b) The PageRank score of each node is shown under $PR$. Nodes in each top-k\% of the rank are grouped based on the unique PageRank scores. In this example, the top-60\% of nodes concentrate most of the PageRank and their scores are somewhat similar (i.e., low Gini). Also, the ranking is fair from top-80\% onwards, since they capture the same fraction of minorities as in the population, 25\%. 
    Local values are measured per top-k\%, and global values are measured using the whole distribution for inequality (Gini), and the average across all top-k\% ranks for inequity (mean error).}
    \label{fig:DBAH}
\end{figure}

\subsection*{Growth network model with homophily and directed links}

In order to examine the effect of homophily on the ranking of minorities in social networks, first we need to develop realistic network models that capture not only a variety of group mixing, but also the directionality of links. Many online social networks are directed networks in their nature, including the follower-followee structure on Twitter, citation networks \cite{kong2021first}, and the hyperlink structure of the Web. Directed links are the key components of many algorithms such as Google Scholar \cite{rovira2019ranking}, PageRank and Who-to-Follow. 

To this end, we propose {DPAH}, a \textbf{d}irected \textbf{p}referential \textbf{a}ttachment with \textbf{h}omophily network growth model. We generate these networks by adjusting the number of nodes $n=2000$, the edge density $d=0.0015$, the fraction of minorities $f_m \in \{0.1, 0.2, 0.3, 0.4, 0.5\}$, the in-class homophily $h_{MM}, h_{mm} \in \{0.0, 0.1, ..., 1.0\}$, and the power-law exponents of the activity distributions $\gamma_M = \gamma_m = 3.0$. We refer to the minority group as $m$, and to the majority group as $M$. Note that the between-class homophily is the complement of the in-class homophily. That is, $h_{Mm}=1-h_{MM}$ and $h_{mM}=1-h_{mm}$.
Furthermore, an activity score is assigned to every node. This score is drawn from a power-law distribution and determines with what probability the existing node becomes active to create additional links to other nodes. This means that more active nodes possess higher out-degree (see Methods for more details).
Each combination of network structure is generated $10$ times, nodes are ranked using PageRank and WTF separately, and inequality and inequity scores are computed and averaged across network types (and top-k's for local disparity) for each algorithm.

\Cref{fig:DBAH} (a) illustrates the generation of a network using the {DPAH}~model. First, $n$ labeled nodes are created. In this example $n=8$. Then, at time $t$, node $p$ is selected as source node with a probability proportional to its activity. Then, $p$ connects to an existing node $j$ with a probability related to their pair-wise homophily $h_{pj}$ and preferential attachment that is based on $k^{in}_{j}$, the in-degree of node $j$. By this process, we ensure that the out- and in-degree distributions of nodes follow seemingly power-law distributions that have been observed in many large social networks \cite{voitalov2019scale}. The algorithm stops once the network reaches an expected density. %
Note that source nodes can be either new nodes joining the network for the first time (e.g., node $p$ at time $t$) or existing nodes (e.g., node $l$ at time $t+1$). Since the network size is given, ``a node joining the network for the first time'' is a 0-degree node that has been selected to create its first edge. Once a source node connects to a target node successfully, the source node becomes available in the next rounds to become a target candidate. This means that in the beginning the model faces a cold start problem since there are no existing (target) nodes to connect to. Thus, the first 1\% of new edges are between a source node (drawn from the activity distribution) and any other node with probability as in \Cref{eq:PAH}. For the sake of completeness we show the computation of local and global disparities of this network in \Cref{fig:DBAH} (b). 

\subsection*{How do homophily and directional links influence the ranking of minorities globally and locally?}

\paragraph{Global disparity.} %

As expected, we found that the Gini coefficient of the rank distributions is large.
$Gini_\text{global}\geq 0.6$ (regions I, II and III; dark colors) for both PageRank (see \Cref{fig:vh_pagerank_DBAH}) and WTF (see Supplementary Figure S1). 
As we will see later, this is mainly due to the preferential attachment mechanism~\cite{pandurangan2002using, fortunato2007local}.
Moreover, we find that on average: (i) Balanced networks ($f_m=0.5$) can get a fair ranking (green) when both groups possess the same homophily scores ($h_{MM}=h_{mm}$). The same applies for neutral networks ($h_{MM}=h_{mm}=0.5$) regardless of their fraction of minorities ($f_m\leq0.5$).
(ii) When the fraction of minorities decreases ($f_m<0.5$), groups can be fairly represented in the rank in two regimes: First, when both groups are homophilic, homophily within minorities must be higher than homophily within majorities ($h_{mm}>h_{MM}>0.5$). Second, when both groups are heterophilic, homophily within majorities must be higher than homophily within minorities ($h_{mm}<h_{MM}<0.5$) to balance the importance of groups.

\begin{figure}[t!]
    \centering
    \includegraphics[width=\textwidth]{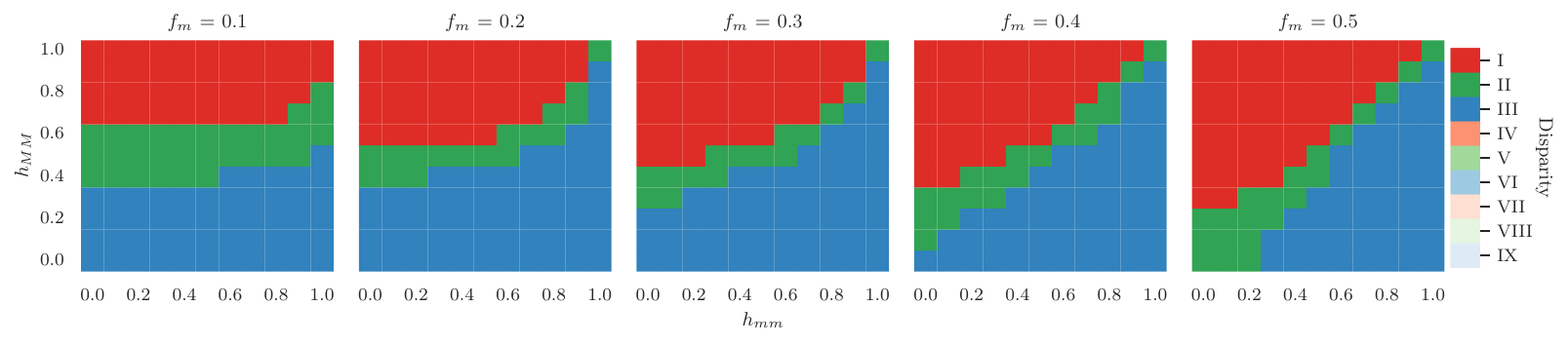}
    \caption{\textbf{The effects of homophily and fraction of minorities in the global disparity of PageRank.} Columns represent the fraction of minorities in the network, x-axis indicates the homophily within minorities, and y-axis the homophily within majorities. 
    Colors denote the region where the disparity lies in according to our interpretation in Figure \ref{fig:interpretation}.
    First, we see that, on average, there is never low global inequality (i.e., regions IV to IX ---lighter colors--- do not appear). This makes sense because these are scale-free networks. Second, depending on the level of homophily within groups, minorities on average can be under-represented (region I, red), or over-represented (region III, blue), or well-represented (region II, green). For example, when $f_{m}=0.1$, minorities are on average under-represented when $h_{MM}\geq 0.7$ and $h_{MM}\geq h_{mm}$.}
    \label{fig:vh_pagerank_DBAH}
\end{figure}

\begin{figure}[t!]
    \centering
    \includegraphics[width=0.7\textwidth]{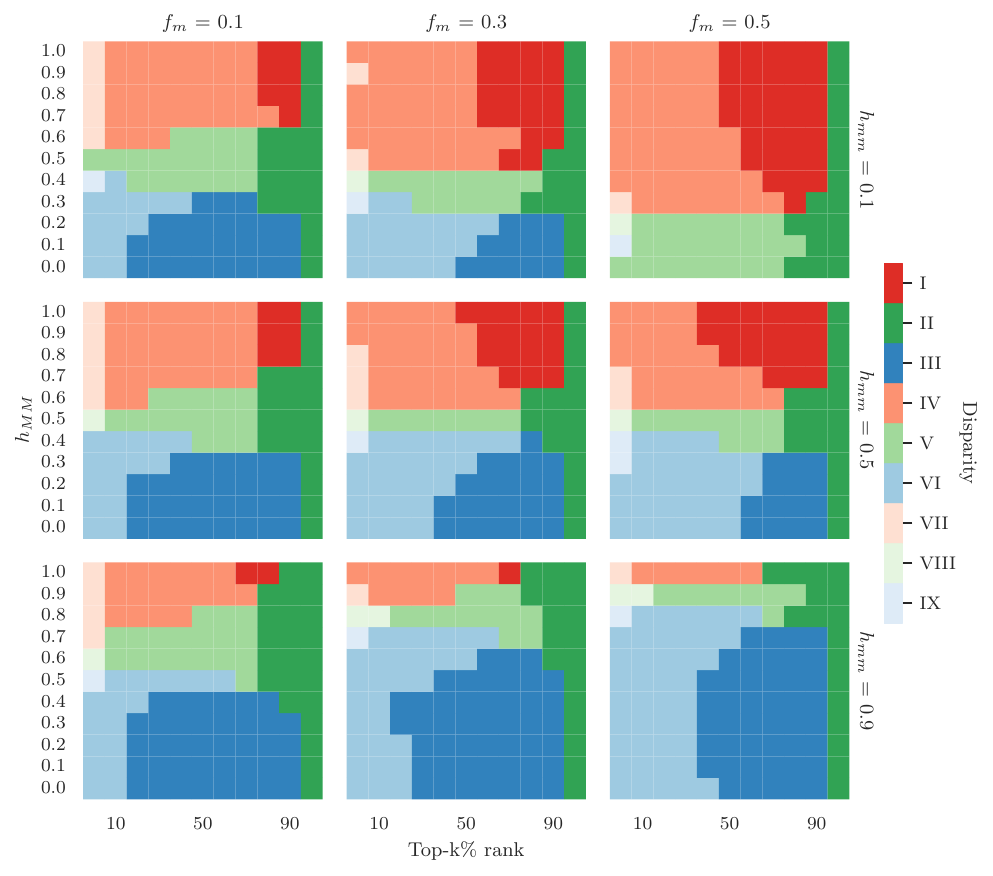}
    \caption{\textbf{The effects of homophily and fraction of minorities in the local disparity of PageRank.} Columns represent the fraction of minorities (10\%, 30\% and 50\%) and rows show homophily within minorities (from top to bottom: heterophilic, neutral and homophilic). 
    The x-axis denotes the top-k\% rank and the y-axis shows homophily within majorities. 
    Colors refer to the regions of disparity introduced in \Cref{fig:interpretation}. 
    One can see that the minority suffers most (red) when the majority is homophilic and the minority is either heterophilic or neutral. 
    Moreover, inequality is lowest (very light colors) only for a few cases at top-5\%. 
    This means that the top best ranked nodes are very similar and their ranks are far from the majority of nodes (i.e., due to preferential attachment). 
    Moreover, inequity remains mostly consistent regardless of top-$k$\%. 
    In other words, if the ranking algorithm favors one group in the top-5\% (e.g., red or blue), it will continue to do so until entering the fair regime (green).}
    \label{fig:h_pagerank}
\end{figure}

\paragraph{Local disparity.}%
We also compute inequality and inequity within each top-k\% rank in order to see to what extent they change when $k$ increases. %
In the case of PageRank, we see in \Cref{fig:h_pagerank} that inequality varies (i.e., from light to dark colors) in different regimes mainly due to the size of $k$ (x-axis), and inequity due to the interplay between homophily within groups, $h_{MM}$ and $h_{mm}$. 
In particular: (i) Only at the top-5\% of the rank we see a few cases of low inequality (regions VII, VIII and IX; very light colors), this means that nodes at the very top possess very similar ranking scores, but they are very far from the rest of the population, i.e., the larger the top-k\%, the higher the Gini (darker colors). This holds for WTF up to roughly the top-30\% (see Supplementary Figure S2). Overall, PageRank converges to high inequality faster than WTF.
(ii) Inequity (regions: red, blue, green) is consistent across all top-k\% ranks for both algorithms. In other words, if the ranking algorithm favors or harms one group in the top-5\%, it will continue to do so until converging to the fair regime (regions II, V, VIII; green). With a few exceptions, this fair regime is only reached when $k$ is very large. For example, if a minority group is under-represented at the top-5\%, it will remain under-represented at the top-80\% (see $h_{mm}=0.1$ and $h_{MM}\geq0.7$ in \Cref{fig:h_pagerank} for PageRank, and Supplementary Figure S2 for WTF).
(iii) Minorities are often over-represented when majorities are heterophilic $h_{MM}<0.5$; (regions III, VI, IX; blue). In contrast, minorities are often under-represented when majorities are homophilic $h_{MM}>0.5$ (regions I, IV, VII; red). This is consistent up to $\approx$ top-$80\%$ for both algorithms.

\hl{In summary, our results suggest that the size of $k$ does not have an influence on inequity. This means that if the algorithm amplifies inequity at the top-5\%, it will also amplify inequity at larger top-k\%'s.
Therefore, increasing the selection pool (larger k) does not improve the representation of minorities. %
This can be explained by the fact that the preferential attachment mechanism disproportionately affects nodes ranking} \cite{ghoshal2011ranking}. 

\paragraph{Correlation and feature importance.} 
We compute the Spearman correlation between inequality and inequity, and conduct a random forest regression to %
measure the importance of each network property on both inequality and inequity values (see Supplementary Appendix A.3 for more details). Results are shown in \Cref{tbl:crossval_pagerank} for PageRank and Supplementary Table S2 for WTF. 
We find that inequality and inequity are positively correlated in both global and local regimes. 
In other words, the more skewed the rank distribution (i.e., high Gini), the more unfair with either group (i.e., mean error far from zero), and vice versa. This correlation is stronger and more significant in PageRank than in WTF. %
In terms of feature importance, we find that global inequality ($Gini$) is mainly explained by both homophily values, whereas global inequity ($ME$) is mainly driven by homophily within majorities. Local inequality ($Gini_k$), however, is mainly explained by the top-$k\%$ rank, and local inequity ($ME_k$) by the homophily within the majority group. %
Notice that we added the variable $\epsilon$ to verify whether the network-based features are better than random (see Supplementary Appendix A.3 for more details). In the case of PageRank, all network-based features perform better than by chance. However, randomness seems to be more relevant for explaining rank inequality ($Gini$) in WTF.

\begin{table}[t!]
\centering
\caption{\textbf{10-fold cross-validation for PageRank.} We use a Random Forest Regressor to assess feature importance and report the mean and standard deviation of the out-of-sample $R^2$. Features are ranked in descending order based on their mean importance (from left to right) and highlighted if their importance represents at least 50\% of the total importance. Corr shows the Spearman correlation between inequality and inequity scores (p-values $\approx0$). \hl{$\epsilon$ represents random chance.}}
\label{tbl:crossval_pagerank}
\begin{tabular}{@{}llllll@{}}
\toprule
\textbf{Type} & \textbf{Outcome} & \textbf{Corr} & \multicolumn{1}{c}{\textbf{$R^2$}} & \textbf{Feature} & \textbf{Importance} \\ \midrule
\multirow{2}{*}{\textbf{Global}} & $Gini$ & \multirow{2}{*}{0.41} & 0.91 (0.009) & $\pmb{h_{MM}}, \pmb{h_{mm}}, f_m, \epsilon$  & 0.43, 0.31, 0.21, 0.05\\
 & $ME$ & & 0.99 (0.001) & $\pmb{h_{MM}}, h_{mm}, f_m, \epsilon$ & 0.61, 0.31, 0.08, 0.0 \\
\multirow{2}{*}{\textbf{Local}} & $Gini_{k}$ & \multirow{2}{*}{0.21} & 0.95 (0.002) &  $\pmb{k}, h_{MM}, h_{mm}, f_m, \epsilon$ & 0.73, 0.11, 0.07, 0.06, 0.03 \\
 & $ME_{k}$ & & 0.99 (0.001)&  $\pmb{h_{MM}}, h_{mm}, k, f_m, \epsilon$ & 0.51, 0.27, 0.14, 0.08, 0.01 \\ \bottomrule
\end{tabular}
\end{table}

These results are in agreement with what we see in previous figures; even though majority nodes produce most of the inequality and inequity in the rank, their interplay with minority nodes can change or intensify the direction of bias. In fact, both homophily values can explain 
$75\%$ (49\%) of $Gini$, the global inequality in PageRank (WTF), 
$92\%$ (88\%) of $ME$, the global inequity, and 
$78\%$ (74\%) of $ME_k$, the local inequity. 
However, the top-k\% rank together with the homophily within majority nodes explain $84\%$ (86\%) of $Gini_k$, the local inequality.

\subsection*{How do different social mechanisms of edge formation contribute to disparity?} %

So far, we show that PageRank and WTF on our network model produce high inequality and a wide-range of possible inequity outcomes. 
How much of that inequality or inequity was a product of homophily or preferential attachment? 
To see the effects of these two mechanisms alone, we generate new networks by turning on and off the homophily and preferential attachment features (see Methods for the details of the models).  %

\begin{figure}[ht]
    \centering
    \includegraphics[width=0.7\textwidth]{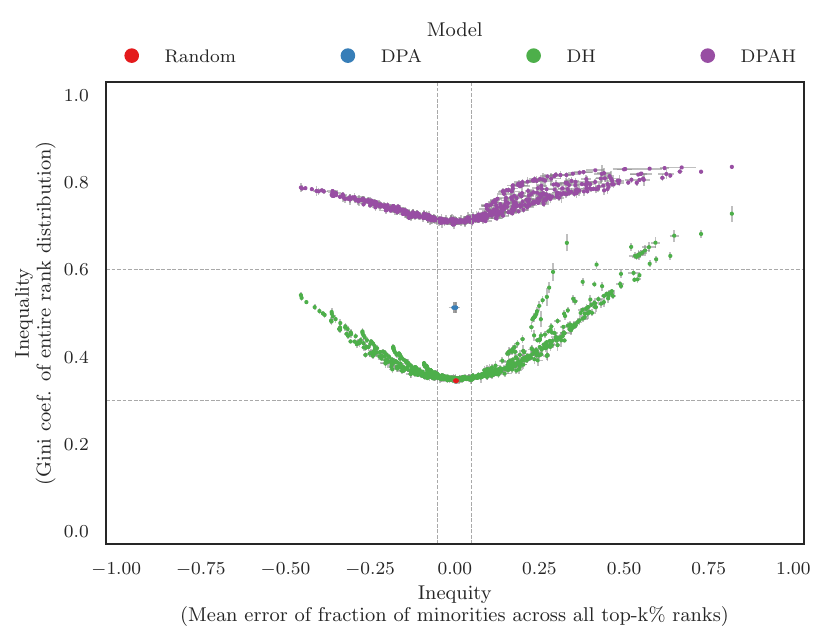}
    \caption{\textbf{The effects of homophily and preferential attachment in the global disparity of PageRank.} We generated directed networks using four different models of edge formation. DPA: only preferential attachment. DH: only homophily. {DPAH}: our proposed model that combines DPA and DH. Random: a baseline where nodes are connected randomly. 
    We see the following patterns: 
    (i) Homophily (DH) produces a moderate-to-high level of inequality ($0.3<Gini<0.8$), while preferential attachment (DPA) produces a consistent moderate inequality ($Gini\approx0.5$). When both mechanisms are combined (DPAH), the rank inequality increases even further ($0.7<Gini<0.9$).
    (ii) Random and Preferential attachment (DPA) are always fair ($ME=0$ or $|ME|\leq \beta$), while in the cases where homophily is involved (DH and DPAH) inequity is often high ($|ME|>\beta$).
    Thus, in general preferential attachment is the main driver of inequality, while homophily influences both inequality and inequity. 
    \hl{Vertical and horizontal error bars represent the standard deviation over 10 runs of the Gini and ME, respectively.}}
    \label{fig:vh_pagerank}
\end{figure}

\Cref{fig:vh_pagerank} shows the inequality and inequity produced by PageRank on a variety of models: DPA (Directed Preferential Attachment), DH (Directed Homophily), Random, and {DPAH}~(see Supplementary Figure S3 for WTF). 
Results from both algorithms show that networks whose nodes connect through preferential attachment (DPA) produce on average higher inequality compared to DH and Random. However, when preferential attachment is combined with homophily ({DPAH}), this inequality increases even further. Additionally, we see that WTF produces higher inequality compared to PageRank (see Supplementary Appendix A.4 for more details). Inequity, on the other hand, is mainly driven by homophily. This means that, homophily ({DPAH}~and DH) influences both, inequality and inequity in both algorithms.

Note that in \Cref{fig:vh_pagerank}, we fixed the activity of nodes to $\gamma_M=\gamma_m=3.0$. 
However, when we set these parameters to $\gamma_M=\gamma_m<3.0$ (more active nodes or lower values of $\gamma$ as found in several scale-free networks~\cite{albert2002statistical}), inequality decreases, see Supplementary Figure S5. 
This behavior holds even if the minority group is the only one increasing its activity ($\gamma_m=1.5 < \gamma_M=3.0$) which in turn increases inequity against the majority, see Supplementary Figure S6.
Additionally, in Supplementary Figure S4, we see that edge density also plays a role in the inequality produced by PageRank and WTF. 
This means that, by further adjusting these two parameters (node activity and edge density), we would expect changes only to inequality since inequity is mainly affected by homophily as we saw before.

\subsection*{Disparities on empirical networks}
\begin{figure}[ht]
    \centering
    \includegraphics[width=1\textwidth]{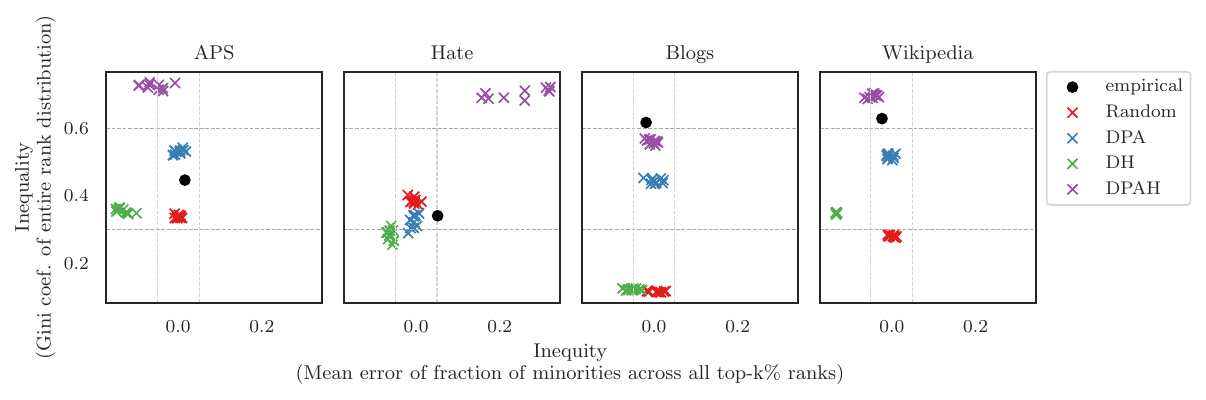}
    \caption{\textbf{Global disparity in PageRank on empirical networks.} Each column represents an empirical network. Citation/retweet networks (APS and Hate) and Hyper-link networks (Blogs and Wikipedia).
    Inequality and inequity are shown in the y- and x-axis, respectively. The disparity in ranking that we see in empirical networks are best explained as follows: (i) citation/retweet networks by preferential attachment PA, and (ii) hyper-link networks by preferential attachment and homophily {DPAH}.}
    \label{fig:empirical}
\end{figure}

First, we fit the DPA, DH, and DPAH models to each of the empirical networks in order to find the mechanism that best explains the inequality and inequity found in the rank. The parameters passed to these models are inferred from the real networks and described in \Cref{tbl:empirical}.
Second, we rank nodes in the empirical and fitted networks using PageRank and WTF, and compute the disparities (inequality vs. inequity) found in their rank distribution. Results are shown in \Cref{fig:empirical} for PageRank and Supplementary Figure S9 for WTF. Disparity values from the real-world networks are labeled as \textit{empirical} (black dot), and disparity values from the fitted networks are labeled according to the model (x marks). 
We see that each network tells a different story.
This can be explained by the nature or domain of these networks. For instance, APS and Hate are best explained by the DPA model. This means that scientists tend to cite authors that have already many citations, and users in Twitter tend to retweet content posted by popular users (i.e., popular in terms of the number of retweets they get). Blogs and Wikipedia on the other hand, are best explained by our {DPAH}~model. Notice that both are hyper-link networks. In other words, people tend to add not only popular references to their Web pages, but also related to their topics (i.e., political leaning in Blogs, and gender in Wikipedia). 
Note that the Hate network shows the lowest (empirical) inequality. This is due to the fact that it possesses low out-degree exponents ($\gamma_M=2.2$, $\gamma_m=1.7$). 

\subsection*{Strategies towards a fair ranking}
Results from both algorithms show that while the homophily within majorities is the main driver for inequality and inequity, minorities may overcome unfair rankings by connecting strategically in the network.
For instance, when both groups are equally active, minorities should adjust their homophily based on the homophily of the majority. 
(i) When majorities are homophilic $h_{MM}>0.5$, minorities should increase their homophily such that $h_{mm}>h_{MM}$. 
(ii) When majorities are (somewhat) neutral ($h_{MM}=0.5\pm 0.1$), minorities may connect arbitrarily with any group without being too homophilic, otherwise they will become over-represented in the rank. 
(iii) When majorities are heterophilic $h_{MM}<0.5$, one solution to achieve a fair rank is to increase the size of the minority group, and make sure that both groups behave similarly in terms of homophily ($h_{MM}\approx h_{mm}$). 
Otherwise, minorities will be over-represented regardless of their in-class homophily.
\hl{On the other hand, when one group is more active than the other, achieving a fair rank becomes challenging. Nevertheless, if the objective is to increase the visibility of minorities in the rank, then the minorities themselves should be more active in the network by creating more connections to increase their out-degree.}
Note that these ``strategies'' without algorithmic intervention may work in scenarios such as a citation or collaboration networks, but they might not work in other scenarios. In such cases, we need additional recommender systems to help under-represented groups discover those ``strategic'' links that will help them climb to higher ranks.

\section*{Discussion and Future Work}
In this work we have proposed a systematic study to measure the inequality and inequity produced by PageRank and Who-To-Follow (WTF). Our approach disentangles the effect of network structure on the rank distributions of these two algorithms by using synthetic networks. By doing so, we control for the properties of the network and measure how these changes affect the rankings. In particular, we studied six prominent structural properties of social networks: homophily, preferential attachment, fraction of minorities, edge density, node activity and the directionality of links. We found that the systemic bias produced by these algorithms in the rank is mainly due to \textit{homophily imbalance} ($h_{MM}\gg h_{mm}$ or $h_{mm}\gg h_{MM}$) for inequity, and the interplay between our six properties of interest for inequality.
Consequently, our systematic study makes PageRank and Who-To-Follow interpretable and explainable since our results show the necessary structural conditions to achieve a fair rank.
A potential avenue to reduce inequity is to then create synthetic connections before the ranking as it is done for correcting the class imbalance problem in supervised learning \cite{chawla2002smote}.
Alternatively, these conditions or strategic connections may be added into the network to change its structure as a collective fairness intervention. 
For instance, recommender systems could suggest relevant articles not only based on popularity and (keyword) similarity but also based on fairness by fulfilling diversity constraints. 

\hl{Notice that our model simplifies the role of homophily and minorities. First, it assumes that all nodes of the same group have the same in-class and between-class homophily. This means that rich mixing patterns might get ignored since some nodes can exhibit local differences} \cite{peel2018multiscale}.
\hl{Second, while there exist multiple definitions of minorities}~\cite{smith1987some} \hl{we adopted the one by Italian jurist Francesco Capotorti: ``\textit{a group numerically inferior to the rest of the population of a State, in a non-dominant position...}''}~\cite{capotorti1979study, hannum2007concept}. \hl{However, we constrained this population to all nodes in a given network (neither the State nor the world population). 
Notice as well that the ranking amplifies the representation of minorities by reducing the representation of majorities in top ranks (e.g., when the majority is heterophilic, see} \Cref{fig:example}). \hl{This aligns with the definition of minorities by Wirth}~\cite{wirth1945problem} \hl{that implies that ``\textit{minorities objectively occupy disadvantageous positions in society}''. This means that ``\textit{a minority may actually, from a numerical standpoint, be the majority}'' (e.g., people living in poverty in under-developed countries). 
In other words, under these two definitions, being in disadvantage in the top-k\% (inequity) is not a group size issue only, but a combination of group size and homophily as we have previously shown.
More complex definitions of minorities are out of the scope of this paper.
A further limitation is that we focus on a single binary attribute (e.g., $\text{color}\in \{\text{Black}, \text{White}\}$), this means that multiple sources of inequality and inequity (e.g., intersections of disadvantage such as being poor and of color) cannot be captured at once}~\cite{hurtado2018intersectional}. \hl{Addressing these issues is beyond the scope of this paper and we leave them for future work.}

Finally, we disentangled the individual effects of preferential attachment and homophily in the rank by comparing the disparities of our proposed {DPAH}~model with two variants and a baseline: networks with preferential attachment only (DPA), networks with homophily only (DH), and directed Erd\"{o}s-R\'{e}nyi (Random)~\cite{erdos1959random} graphs.
Further research can investigate other topologies and social mechanisms of edge formation %
such as clustering \cite{davis1970clustering}, transitivity \cite{block2015reciprocity}, and reciprocity \cite{dufwenberg2017reciprocity}. Similarly, other structural properties such as monophily \cite{altenburger2018monophily} and second order homophily \cite{Evtushenko2021paradox} can be studied to measure their influence on ranking.

\section*{Conclusions}
In this work we have investigated under which conditions {PageRank} and Who-To-Follow (WTF) \hl{\textit{reduce}, \textit{replicate} or \textit{amplify} the representation of minorities in top ranks. }
In particular, given the rank distribution produced by these algorithms, we computed \textit{inequality} as the dispersion among individuals in terms of ranking scores, and \textit{inequity} as whether minorities are over-, under- or well-represented in top ranks compared to their representation in the network. %
We studied these two metrics separately and in combination to better understand the mechanisms that can explain them. 

To that end, we proposed {DPAH}, a growth network model that allows to generate realistic scale-free directed networks with different levels of homophily, fraction of minorities, node activity, and edge density. 
In these networks, we found that both inequality and inequity are positively correlated and mainly driven by the homophily within majorities. 
This means that, when the majority group is highly homophilic, the minority group is under-represented in top ranks. Also, when the majority is highly heterophilic, the minority benefits tremendously since it is over-represented in the top-k\%. However, minorities can overcome these disparities by connecting strategically with others. Thus, equity in ranking is a trade-off between homophily and the fraction of minorities. 

Our systematic study makes PageRank and Who-to-Follow explainable and interpretable to help data scientists understand and estimate the disparity that these algorithms produce given the structure of networks, which is key for proposing targeted interventions.
We hope our results create awareness among majority and minority groups about these disparities %
since they may replicate and even amplify the biases found in social networks.

\section*{Data and Methods}

\subsection*{Synthetic networks}
Network models have been proposed with various social mechanisms. For instance, the classic \textit{stochastic-block model} \cite{holland1983stochastic} which allows for homophily between and across groups, and the \textit{configuration model} \cite{newman2003structure} which generates links among nodes by preserving a given degree distribution. 
On the other hand, the \textit{preferential-attachment} model \cite{barabasi1999emergence} produces scale-free networks due to cumulative advantage \cite{merton1988matthew}. 
Although these models can reproduce certain properties of real-world networks such as degree or homophily, they fail at guaranteeing similar visibility of minorities as their empirical counterpart. 
In this direction, Karimi et al.\cite{karimi2018homophily} and Fabbri et al.~\cite{fabbri2020effect} devise social network models with preferential attachment, adjustable homophily and fraction of minorities. 
They demonstrate how the degree rank of the minority group in a network is a function of the relative group sizes and the presence or absence of homophily.
However, the former models undirected networks, and the latter did not control for edge density and node activity (i.e., power-law out-degree distributions) as we do in this work for minority and majority groups.

\subsubsection*{Directed network}
We define a directed network as: Let $G=(V,E,C)$ be a node-attributed unweighted graph with $V=\{v_1,...,v_n\}$ being a set of $n$ nodes, $E \subseteq  V \times V$ a set of $e$ directed edges, and $C=\{c_1,...,c_n\}$ a list of binary class labels where each element $c_i$ represents the class membership of node $v_i$. The fraction of minorities $f_m$ captures the relative size of the minority class---with respect to $C$---in the network. We refer to the minority group as $m$, and to the majority group as $M$. A network is \emph{balanced} when all class labels have the same number of nodes ($f_m=0.5$), otherwise it is \emph{unbalanced} ($f_m<0.5$). \hl{Networks fulfill a predefined edge density level $d$. Since $n$ and $d$ are given, networks stop growing when $e=d n (n-1)$.} 

In order to generate directed links, inspired by the activity-driven network model \cite{perra2012activity}, we assign an activity score to each node that determines with what probability the existing node becomes active and creates additional links to other nodes. It has been shown that in empirical networks the activity of the nodes follows a power-law distribution \cite{perra2012activity}. Therefore, we assign an activity to each node drawn from a power-law distribution \hl{$\rho(\gamma) = X^{-\gamma}$}. Note that each group possess its own activity distribution and they are defined by its power-law exponent $\gamma_M$ and $\gamma_m$ for majority and minority nodes, respectively. \hl{The level of activity of a group is inversely proportional to $\gamma$. That is, groups with higher out-degree produce lower $\gamma$ (more skewed).}

Then, the probability of connecting a source (active) node $v_i$ to a target node $v_j$ (or in other words the probability of connecting to $v_j$ given the source node $v_i$) is explained by any of the following three mechanisms of edge formation.

\subsubsection*{Preferential Attachment (DPA)}
Also known as the \textit{rich-get-richer} effect or \textit{cumulative advantage} in social networks \cite{merton1988matthew, barabasi1999emergence}. It indicates that nodes tend to connect to popular nodes. We define popularity as the in-degree of the node. Therefore, the probability that a source node $v_i$ connects to a target node $v_j$ is proportional to the \textit{in-degree} of the target node $v_j$. 
\begin{equation}
    P(i\to j) = P(j|i) = \frac{k^{in}_{j}}{\sum_{l=1}^{N} k^{in}_{l}}
\label{eq:PA}    
\end{equation}

\subsubsection*{Homophily (DH)}
It is the tendency of individuals to connect (or interact) with similar others~\cite{mcpherson2001birds,newman2003structure}. 
Thus, the probability that a source node $v_i$ connects to a target node $v_j$ is driven by the homophily between their classes $c_i$ and $c_j$. 
We assign a homophily value to each dyad based on pre-defined homophily parameters within majorities and minorities, $h_{MM}$ and $h_{mm}$, respectively. Homophily values range from $0.0$ to $1.0$. If the homophily value is high, that means that nodes of the same class are attracted to each other more often than nodes of different attributes. 
Following the definitions from previous work~\cite{rogers1970homophily, karimi2018homophily, fabbri2020effect}, nodes of the same class with homophily $h_{aa}=0.5$ are referred to as \emph{neutral} (i.e., they connect randomly to either class), otherwise they are \emph{heterophilic} if $h_{aa}<0.5$ (i.e., more likely to connect to the other class), or \emph{homophilic} when $h_{aa}>0.5$ (i.e., more likely to connect to the same class). Note that in- and between-class homophily values are complementary: $h_{mm}=1-h_{mM}$ and $h_{MM}=1-h_{Mm}$. 
\begin{equation}
    P(i\to j) = P(j|i) = \frac{h_{ij}}{\sum_{l=1}^{N} h_{il}}
    \label{eq:H} 
\end{equation}

\subsubsection*{Preferential Attachment with Homophily ({DPAH})}
We propose {DPAH}, a directed growth network model with adjustable homophily and fraction of minorities. {DPAH}~stands for \textbf{D}irected network with \textbf{P}referential \textbf{A}ttachment and \textbf{H}omophily. This mechanism combines DPA and DH, and is an extension of the BA-Homophily model \cite{karimi2018homophily}.

\begin{equation}
    P(i \to j) = P(j|i) = \frac{h_{ij}k^{in}_{j}}{\sum_{l=1}^{N} h_{il} k^{in}_{l}}
    \label{eq:PAH} 
\end{equation}

Note that DPA and DH are especial cases of {DPAH}~where only the in-degree mechanism varies. This means that, the out-degree distribution remains the same as in {DPAH}: it is driven by the activity model. 
Additionally, we include a random model where both source and target nodes are chosen at random (i.e., directed Erd\"{o}s-R\'{e}nyi  model \cite{erdos1959random}).
\Cref{tbl:synthetic} shows the parameters adjusted in each model. Number of nodes $n$ and edge density $d$ are arbitrary in the sense that they are not part of the edge formation mechanism. Thus, we fix them to make a fair comparison across all models.

\begin{table}[ht]
\centering
\caption{\textbf{Model parameters.} 
Check marks denote that a given model (column) requires a particular parameter (row): number of nodes $n$, fraction of minorities $f_m$, edge density $d$, in-class homophily $h_{aa}$, and the power-law exponent of the activity distribution $\gamma$. Sub-indices $M$ and $m$ refer to the majority and minority groups, respectively.
The difference between DH and DPAH is the preferential attachment (in-degree) mechanism. All models produce directed networks.}
\label{tbl:synthetic}
\begin{tabular}{@{}ccccc@{}}
\toprule
 & \textbf{Random} & \textbf{DPA} & \textbf{DH} & \textbf{DPAH} \\ \midrule
\textbf{$n$} & \checkmark & \checkmark & \checkmark & \checkmark \\
\textbf{$f_m$} & \checkmark & \checkmark & \checkmark & \checkmark \\
\textbf{$d$} & \checkmark & \checkmark & \checkmark & \checkmark \\
\textbf{$h_{MM}$} & - & - & \checkmark & \checkmark \\
\textbf{$h_{mm}$} & - & - & \checkmark & \checkmark \\
\textbf{$\gamma_M$} & - & \checkmark & \checkmark & \checkmark \\
\textbf{$\gamma_m$} & - & \checkmark & \checkmark & \checkmark \\ \bottomrule
\end{tabular}
\end{table}

\subsection*{Empirical networks}
 We inspect four networks from different domains and compute the inequalities and inequities produced by PageRank and WTF. \Cref{tbl:empirical} shows the most important properties of these networks.
\begin{itemize}
    \item \textbf{APS:} The American Physical Society citation network whose nodes represent articles, and edges represent citations. The binary class of each node is $\text{pacs}$ and encodes two different Physics sub-fields where $\text{05.20.-y}$ (Classical statistical mechanics) is the minority.

    \item \textbf{Hate:} A retweet network \cite{kaggle:hate} where nodes denote users, and edges represent retweets among them. Users are labeled as either $\text{hateful}$ or $\text{normal}$ depending on the sentiment of their tweets. Hateful users represent the minority.
    
    \item \textbf{Blogs:} An hyper-link network from political blog posts about the 2004 U.S. election \cite{adamic2005political}. Nodes represent blog pages, and edges hyper-links among them. Each blog is labeled as either $\text{right-}$ or $\text{left-}$leaning. The latter represents the minorities.

    \item \textbf{Wikipedia:} A Wikipedia hyper-link network where nodes represent U.S. politicians \cite{wikiwag, wiki:poli} labeled as either male or female. Female politicians represent the minorities. 
\end{itemize}

\begin{table}[ht]
\caption{Empirical Networks. APS, a scientific citation network. Hate, a retweet network. Blogs, a political blog hyper-link network. Wikipedia, a hyper-link network of politicians. Each row represents a property of the network. $E_{**}$ represents the fraction of edges within and across groups, and $h_{**}$ homophily values inferred by the {DPAH}~model (see Supplementary Appendix A for derivations).}
\label{tbl:empirical}
\centering
\begin{tabular}{llllll}
\toprule
dataset &          APS &     Hate   &  Blogs   & Wikipedia \\
\midrule
$n$     &         1853 &     4971    &   1224  & 3159   \\
$\text{class}$ &  pacs &     hate    & leaning & gender \\
$M$        &  05.30.-d &     normal  &  right  & male    \\
$m$        &  05.20.-y &     hateful &  left   & female  \\
$f_m$      &   0.37561 &     0.10943 & 0.48039 & 0.15226 \\
$d$        &   0.00106 &     0.00061 & 0.01271 & 0.00149 \\
$\gamma_M$ &   3.22246 &     2.23026 & 4.88733 & 4.22425 \\
$\gamma_m$ &   8.93993 &     1.73445 & 3.22464 & 6.16567 \\
$E_{MM}$   &   0.64981 &     0.56898 & 0.47070 & 0.78469 \\
$E_{Mm}$   &   0.02859 &     0.10244 & 0.04741 & 0.07824 \\
$E_{mM}$   &   0.02721 &     0.07886 & 0.04105 & 0.10685 \\
$E_{mm}$   &   0.29439 &     0.24972 & 0.44084 & 0.03022 \\
$h_{MM}$   &   0.94000 &     0.58000 & 0.92000 & 0.59000 \\
$h_{mm}$   &   0.96000 &     0.95000 & 0.90000 & 0.62000 \\
\bottomrule
\end{tabular}
\end{table}

\subsection*{Ranking and Recommendation algorithms}
There exist a variety of ranking and recommendation algorithms that follow different strategies depending on the nature of the problem. For instance, in information systems, items such as content, Web pages, and products are ranked to recommend users what to read or buy \cite{lofgren2016personalized}. In social networks, however, people are ranked to identify their hierarchy or importance \cite{ding2009pagerank, gollapalli2011ranking,senanayake2015pagerank}, and recommended to other users in order to establish new connections \cite{barbieri2014follow, yu2014link, morone2015influence, zhang2016identifying}.
These rankings and recommendations are based on algorithms that often rely on whom we are already connected with.
In this work, we focus on two such algorithms widely used in practice \cite{gleich2015pagerank}: PageRank \cite{page1999pagerank} and Who-to-Follow (WTF) \cite{gupta2013wtf}. While PageRank determines the global ranking of nodes in comparison with all other nodes,  WTF deals with ranking nodes in a node level and thus remains a local measure. For that reason, we focus on these two algorithms to capture both dimensions.

\subsubsection*{PageRank}
It was invented to rank all web pages in the Web \cite{page1999pagerank}, and has been used in several applications \cite{gleich2015pagerank}. For example, to study citation and co-authorship networks \cite{liu2005co, jezek2008exploration, fiala2008pagerank}.
PageRank assigns an importance score to every single node in a network. This score takes into account the number and quality of incoming links of each node. 
The PageRank of node $i$ is defined as follows:
\begin{equation}
    PR(i) = (1-\alpha) + \alpha \sum_{j \in N_i} \frac{PR(j)}{k^{out}_j}
\end{equation}
where $i\in V$, $N_i$ represents all neighbors of node $v_i$ (e.g., all nodes $v_i$ points to), and $k^{out}_j$ the out-degree of node $v_j$. The damping factor $\alpha$, or probability of following links using a Random Walker, is set to $0.85$ as suggested by Brin and Page \cite{brin1998anatomy}. We use the \texttt{fast-pagerank}~\cite{fast-pagerank} python package to compute the PageRank score of all nodes using sparse adjacency matrices.

\subsubsection*{Who-To-Follow (WTF)}
This recommendation algorithm was created and used by Twitter to suggest new people to follow \cite{gupta2013wtf}. It is based on SALSA \cite{lempel2001salsa} which in turn is based on Personalized PageRank \cite{jeh2003scaling}.
In a nutshell, for each user $u$ (or node $v_i \in V$), the algorithm looks for its \textit{circle of trust}, which is the result of an egocentric random walk (similar to personalized PageRank) \cite{gupta2013wtf}. Then, based on this circle-of-trust, the algorithm ranks all users that are not yet friends with $u$ but are connected through the circle of trust. Then, we take the top-k of these (recommended) users, and add up the counter of being selected as a recommendation to each of them. This is done for every node $u$ in the network. At the end, the rank of each node encodes the \textit{number of times a user was suggested as a recommendation} across all nodes in the network. Thus, the WTF score for each node is defined as follows:
\begin{equation}
    WTF(i) = \sum_{j\in V} \mathds{1}_{SALSA(j)}(i)
\end{equation}
where $SALSA(j)$ refers to the top-k users the SALSA algorithm recommends to node $j$. In this work we select the top-$10$ users as recommendations. $\mathds{1}_A(x)$ denotes the indicator function or boolean predicate function to test set inclusion (i.e., whether $x \in A$).

\subsection*{Gini coefficient}
The Gini coefficient was developed by the Italian Statistician Corrado Gini~\cite{gini1912variabilita} to measure the income inequality of a society. 
It is defined as the mean of absolute differences between all pairs of individuals for some measure. In our setup this measure is the score given to every node by PageRank and Who-To-Follow. 
The minimum value is 0 when all individuals' scores are equal, and its maximum value is 1 when there is a big gap or discrepancy between scores \cite{ceriani2012origins}.

We define the Gini coefficient of the rank distribution $X$ as follows. For more details see~\cite{gini}: 

\begin{equation}
Gini(X)=\frac{\sum_{i=1}^{\hat{n}} (2i- \hat{n} -1)x_i}{n\sum_{i=1}^{\hat{n}} x_i}
\end{equation}
 
where $x \in X$ is an observed value in the rank distribution, $\hat{n}=|X|$ is the number of values observed, and $i$ is the rank of values in ascending order.

\section*{Acknowledgements}
Special thanks to Nicola Perra, Indira Sen, Mattia Samori, Reinhard Munz who have helped improving this paper and to Antonio Ferrara and Jan Bachmann for their technical support on the revisions of this paper. 
Also, thanks to LXAI@ICML2020, NetSci2020, and WiDS Guayaquil@ESPOL for their feedback.
This work was partially funded by the Austrian Science Promotion Agency FFG project no. 873927. The presented computational results have been achieved in part using the Vienna Scientific Cluster (VSC).

\section*{Author contributions statement}
L.E.N., F.K., C.W. and M.S. devised the research project. L.E.N. conceived the experiments and analyzed the datasets. F.K. performed analytical derivations. L.E.N., F.K., C.W., and M.S. wrote the paper. All authors reviewed the manuscript. Parts of L.E.N and M.S.'s work on this project have been performed while being at GESIS and RWTH Aachen, respectively.

\section*{Additional information}

\subsection*{Competing Interests}
The author(s) declare no competing interests.

\subsection*{Availability of materials and data}
The code and datasets generated during and/or analyzed during the current study are available in the GitHub repository, \url{https://github.com/gesiscss/Homophilic_Directed_ScaleFree_Networks}.

\clearpage
\setcounter{section}{0}
\setcounter{equation}{0}
\setcounter{table}{0}
\setcounter{figure}{0}
\renewcommand{\thetable}{S\arabic{table}}
\renewcommand{\thefigure}{S\arabic{figure}}

\section*{Supplementary Material}
\appendix
\section{Analytics}
\label{sm:analytics}
\subsection{Derivation of the probability of having an internal link}
\label{sup:probability}

Let $K^{in}_a(t)$ and $K^{out}_a(t)$ be the sum of the in- and out-degrees of nodes from group $a$ at time $t$. The overall growth of the network follows a DPAH process. %
Thus, the evolution of in-degree and out-degree follows:
\begin{equation}\label{req:kakb}
\left\{
  \begin{array}{l}
K_a^{in}(t) + K_b^{in}(t) = K^{in}(t) = m t %
\\
K_a^{out}(t) + K_b^{out}(t) = K^{out}(t) = m t %
  \end{array}
\right.
\end{equation}
where $m$ is the number of new links in the network at each time step $t$. In each time step, a node $v_i$ is chosen. That results in $m$ new out-going links from $v_i$. %
We set $m=1$. Thus, in each time step only one edge is created from $v_i$ to $v_j$.

Let us denote the relative fraction of group size for each group as $f_a$ and $f_b$ \hl{and their respective activity parameters $\gamma_a$ and $\gamma_b$ that represent the exponents of the activity distribution.} Thus, the behavior of the network is similar to what we have shown before \cite{karimi2018homophily}; only the total number of links is different.  
Let us also define $\hat{\gamma_a}$ as an average value drawn from the activity distribution of group $a$, $\rho(\gamma_a) = X^{-\gamma_a}$ using mean field approximation. Similarly, $\hat{\gamma_b}$ as an average value drawn from the activity distribution of group $b$, $\rho(\gamma_b) = X^{-\gamma_b}$.
We can show that in the limit of $\Delta t \rightarrow 0$, for each group, the in-degree growth function follows: %
\begin{equation}
    \label{req:evol_ka}
    \dfrac{dK_a^{in}}{dt} = m \left(f_a \times \hat{\gamma_a} \left(1 + \dfrac{h_{aa}K^{in}_a(t)}{h_{aa}K^{in}_a(t) + h_{ab}K^{in}_b(t)}\right) + f_b  \hat{\gamma_b} \left(\dfrac{h_{ba}K^{in}_a(t)}{h_{bb}K^{in}_b(t) + h_{ba}K^{in}_a(t)}\right)\right)\\
\end{equation}
\begin{equation}
    \label{req:evol_kb}
    \dfrac{dK_b^{in}}{dt} = m \left(f_b \times \hat{\gamma_b} \left(1 + \dfrac{h_{bb}K^{in}_b(t)}{h_{bb}K^{in}_b(t) + h_{ba}K^{in}_a(t)}\right) + f_a \hat{\gamma_a} \left(\dfrac{h_{ab}K^{in}_b(t)}{h_{aa}K^{in}_a(t) + h_{ab}K^{in}_b(t)}\right)\right)\\
\end{equation}

Next, we focus on the case of links within group $a$. The same analysis applies for group $b$. 

Let $p_{aa}$ be the probability to establish a link between two nodes of group $a$. %
The probability for an incoming or existing node from group $a$ to link to a node of the same group is given by:
\begin{equation}
\label{eq:paa}
 p_{aa}(t) = f_a\dfrac{h_{aa}K_{a}^{in}(t)}{h_{aa}K_{a}^{in}(t) + h_{ab}K_{b}^{in}(t)}
\end{equation}

In the simple network growth model, the total degree of the groups increases linearly over time.

\begin{equation}
  \left\{
    \begin{array}{l}
      K_a^{in}(t) = C m \hat{\gamma_a} \hat{\gamma_b} t\\
      K_b^{in}(t) = (2-C) m \hat{\gamma_a} \hat{\gamma_b} t\\
      K_a^{out}(t) = m \hat{\gamma_a} t\\
      K_b^{out}(t) = m \hat{\gamma_b} t\\
    \end{array}
  \right.
\end{equation}
Denoting $C$ as the in-degree growth factor of the minority group.

\subsection{Calculating homophily from empirical network}
\label{sup:homophily}

We can calculate homophily in empirical networks using the information about in-group links. First, the total number of edges in a directed network follows:
\begin{equation}
  e = e_{aa} + e_{ab} +e_{ba}+ e_{bb}
\end{equation}

To calculate $e_{aa}$, the number of links within class $a$, we can simply argue that it depends on $p_{aa}$, the probability of connecting two nodes belonging to class $a$, multiplied by the probability of the arrival or source node to be of class $a$, denoted by $f_a$, the fraction of nodes in class $a$, as shown in \Cref{eq:paa}.

Our network model grows linearly in time. That means, the in-degree growth for each group is linear. Let us assume that the in-degree growth rate of group $a$ is denoted by $C_a$:

\begin{equation}
 K^{in}_a(t) = C_a K^{in}(t)
\end{equation}

Since the in-degree growth remains constant over time, we can calculate $C_a$ in the empirical network by summing all in-degrees of the group
\begin{equation}
 C_a(empirical) = \frac{K^{in}_a}{K^{in}}
\end{equation}

\Cref{eq:paa} can be rewritten as
\begin{equation}\label{req:paa_analytical}
  p_{aa} = f_a\dfrac{h_{aa}C_a}{h_{aa}C_a + h_{ab}(1-C_a)}
\end{equation}

In empirical networks, $p_{aa}$ represents the probability of a directed edge from class $a$ to class $a$. This probability is proportional to the number of edges from $a$ to $a$, normalized by the total number of directed edges that start from $a$:
\begin{equation}\label{req:paa_empirical}
  p_{aa} = \frac{e_{aa}}{e_{aa}+e_{ab}}
\end{equation}

We can then calculate \Cref{req:paa_empirical} in the empirical network. Finally we use maximum-likelihood estimate to find the best values for $h_{aa}$ and $h_{bb}$ in \Cref{req:paa_analytical}.

\hl{Note that the in-degree growth rate $C$ has an sub-linear relationship to the exponent of the in-degree distribution $\sigma$  and the exponent of the in-degree growth $\theta$} \cite{karimi2018homophily}. \hl{Thus, another method to retrieve empirical homophily is to first estimate the exponents of the in-degree distributions for minority and majority groups ($\sigma_a$ and $\sigma_b$) and plug that into the equation. }

\begin{equation}
p_{aa} = \frac{f_a^2 h_{aa}(1-\theta_b)}{f_a h_{aa}(1-\theta_b) + f_b h_{ab}(1-\theta_a)}
\end{equation}

\begin{equation}
p_{bb} = \frac{f_b^2 h_{bb}(1-\theta_a)}{f_b h_{bb}(1-\theta_a) + f_a h_{ba}(1-\theta_b)}
\end{equation}

where $\sigma_a = -(\frac{1}{\theta_a} + 1)$ and $\sigma_b = -(\frac{1}{\theta_b} + 1)$.

\subsection{Regression model}
\label{sm:models}
We build a \texttt{RandomForestRegressor}~\cite{scikit-learn,randomforest} model to explain rank inequality and inequity given the structure of networks. Features (or independent variables) are transformed by scaling them between zero and one. During training, the model uses $n\_estimators=100$ and all default values from the \texttt{Python} package. We use $R^2$ scores to evaluate the performance of the $10$-fold cross-validated model on the test set.
As shown in \Cref{tbl:models}, the global model takes into account the overall behavior or trend regardless of top-k ranks, while the local model includes the top-k ranks. We add $\epsilon$ as a dummy variable, with randomly generated values between 0 and 1, to compare the importance of each feature to random chance.

We report the importance of features given by the \texttt{feature\_importances\_} property of the \texttt{RandomForestRegressor} model. The higher the value, the more important the feature. The importance of a feature is computed as the (normalized) total reduction of the criterion brought by that feature. It is also known as the Gini importance~\cite{randomforest}.

\subsection{Who-To-Follow produces higher inequality compared to PageRank}
\label{sm:pr_wtf}
In the main manuscript we see that Who-To-Follow (WTF) produces skewer rank distributions compared to PageRank.
To understand this behavior, we need to first understand how the algorithms work.
PageRank scores reflect the \textit{global} importance of nodes in the network, and this global importance is mostly determined by in-degree \cite{fortunato2006approximating} and the age of nodes \cite{mariani2015ranking}.
On the other hand, the WTF score of a node is the number of times the node appears in the top-10 recommendation across all nodes in the network. 
This top-10 is determined by the circle-of-trust of each node, similar to a Personalized PageRank.
This means that this top-10 contains the most visited nodes by a random walker that always starts at the node who is getting the recommendation. Thus, that (local) top-10 will be highly influenced by in-degree too.
However, since the WTF score counts the number of times a node appears as a recommendation, it is likely that the highest WTF scores refer to high degree nodes due to preferential attachment.
Therefore, the high inequality produced by WTF can be explained by the fact that WTF combines a local random walk with a global count.

\section{Additional Tables and Figures}

\begin{table}[h!]
\centering
\caption{\textbf{Regression models.} Dependent and independent variables of the four models of interest: Global/Local inequality (Gini) and inequity (ME). We add the dummy variable $\epsilon$ (with randomly generated values between 0 and 1) to verify whether the network-based features are better than random or not.}
\label{tbl:models}
\begin{tabular}{@{}lll@{}}
\toprule
\textbf{Type} & \textbf{Dependent variable (Y)} & \textbf{Independent variable (Xs)} \\ \midrule
\multirow{2}{*}{Global} & $Gini$   & \multirow{2}{*}{$f_m$, $h_{MM}$, $h_{mm}$, $\epsilon$}    \\
                        & $ME$     &                                               \\
\multirow{2}{*}{Local}  & $Gini_k$ & \multirow{2}{*}{$f_m$, $h_{MM}$, $h_{mm}$, $k$, $\epsilon$} \\
                        & $ME_k$   & \\ \bottomrule
\end{tabular}
\end{table}

\begin{table}[h!]
\centering
\caption{\textbf{10-fold cross-validation for WTF.} We use a Random Forest Regressor to assess feature importance and report the mean and standard deviation of the out-of-sample $R^2$. Features are ranked in descending order based on their mean importance (from left to right) and highlighted if their importance represents at least 50\% of the total importance. Features with a mark (*) are less important than random $\epsilon$. Corr shows the disparity as the Spearman correlation between inequality and inequity scores (p-values $\approx0$).}
\label{sm:crossval_wtf}
\begin{tabular}{@{}llllll@{}}
\toprule
\textbf{Type} & \textbf{Outcome} & \textbf{Corr.} & \multicolumn{1}{c}{\textbf{$R^2$}} & \textbf{Feature} & \textbf{Importance} \\ \midrule
\multirow{2}{*}{\textbf{Global}} & $Gini$ & \multirow{2}{*}{0.29} & 0.35 (0.03) & $\epsilon$, $h_{MM}^*$, $h_{mm}^*$, $f_m^*$ & 0.37, 0.27, 0.22, 0.14 \\
 & $ME$ & & 0.92 (0.01) & $\pmb{h_{MM}}, h_{mm}, f_m, \epsilon$ & 0.51, 0.37, 0.07, 0.05 \\
\multirow{2}{*}{\textbf{Local}} & $Gini_{k}$ & \multirow{2}{*}{0.06} & 0.86 (0.00) &  $\pmb{k}, \epsilon, h_{MM}^*, h_{mm}^*, f_m^*$ & 0.86, 0.08, 0.02, 0.02, 0.01 \\
 & $ME_{k}$ & & 0.85 (0.01) &  $\pmb{h_{MM}}, \pmb{h_{mm}}, k, \epsilon, f_m^*$ & 0.43, 0.31, 0.11, 0.08, 0.07 \\ \bottomrule
\end{tabular}
\end{table}

\begin{figure}[h!]
    \centering
    \includegraphics[width=\textwidth]{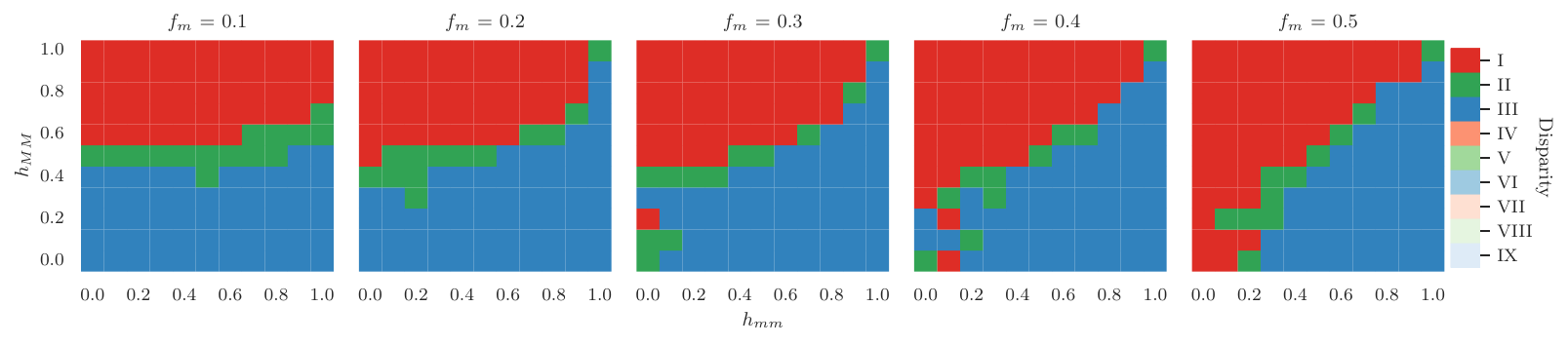}
    \caption{\textbf{The effects of homophily and fraction of minorities in the global disparity of WTF.} Columns represent the fraction of minorities in the network, x-axis indicates the homophily within minorities, and y-axis the homophily within majorities. 
    Colors denote the region where the disparity lies according to our interpretation (see Figure 2 in main article).
    As in the case of PageRank (cf. Figure 4 in main article), we see that, on average, there is never low global inequality. Also, depending on the level of homophily within groups, minorities on average can be under-represented (region I, red), or over-represented (region III, blue). Note that the fair case (region II, green) rarely occurs.
    }
    \label{sm:vh_wtf_DBAH}
\end{figure}

\begin{figure}[h!]
    \centering
    \includegraphics[width=0.6\textwidth]{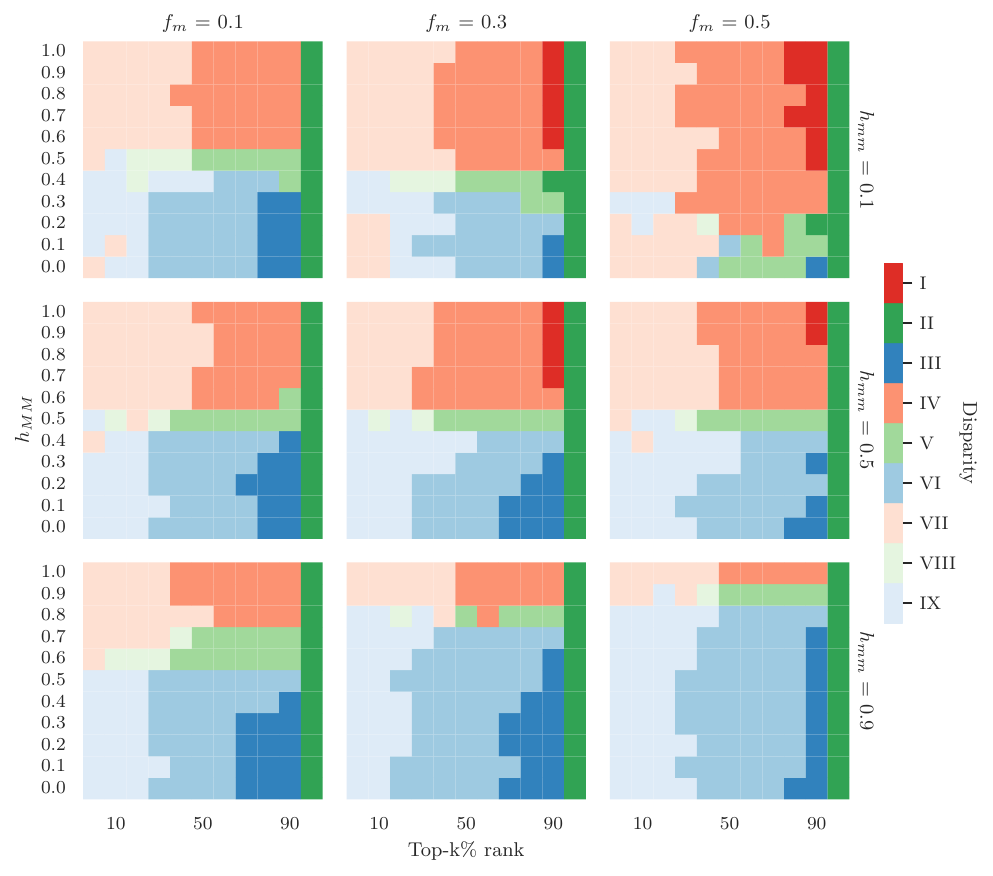}
    \caption{\textbf{The effects of homophily and fraction of minorities in the local disparity of WTF.} Columns represent the fraction of minorities (10\%, 30\% and 50\%) and rows show homophily within minorities (from top to bottom: heterophilic, neutral and homophilic). 
    The x-axis denotes the top-k\% rank and the y-axis shows homophily within majorities. Colors refer to the regions of disparity (see Figure 2 in main article). 
    As in the case of PageRank (cf. Figure 5 in main article), we see that the minority suffers most (red) when the majority is homophilic and the minority is either heterophilic or neutral. 
    Also, inequity remains mostly consistent regardless of top-$k$\%. 
    In contrast to PageRank (up to top-5\%), WTF manages to capture nodes with very similar ranking scores (roughly) up to the top-30\% (i.e., Gini is low, regions VII, VIII, IX).}
    \label{sm:h_wtf}
\end{figure}

\begin{figure}[h!]
    \centering
    \includegraphics[width=0.6\textwidth]{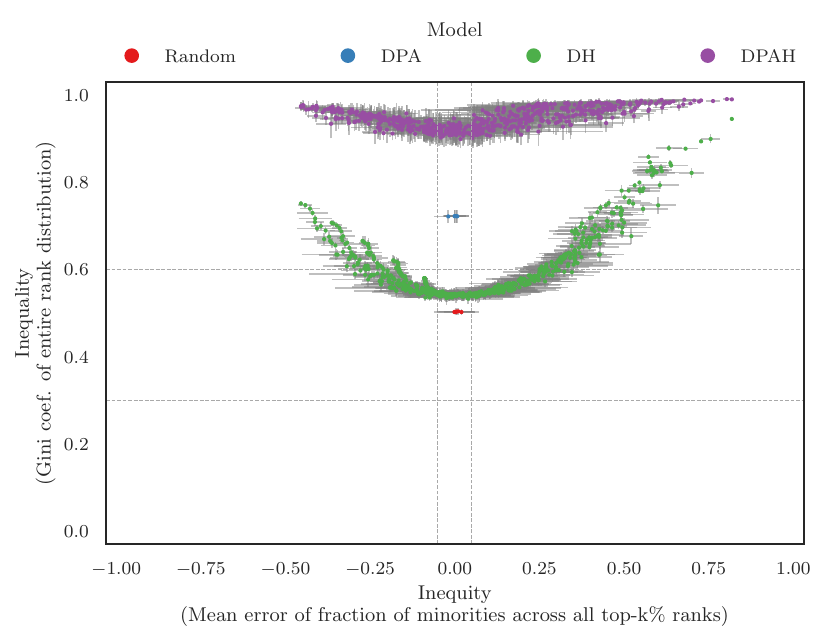}
    \caption{\textbf{The effects of homophily and preferential attachment in the global disparity of WTF.} We generated directed networks using four different models of edge formation. DPA: only preferential attachment. DH: only homophily. {DPAH}: our proposed model that combines DPA and DH. Random: a baseline where nodes are connected randomly. 
    Compared to PageRank (cf. Figure 6 in main article), all models generate higher inequality (y-axis), whereas inequity remains similar. 
    Vertical and horizontal error bars represent the standard deviation over 10 runs of the Gini and ME, respectively.}
    \label{sm:vh_wtf}
\end{figure}

\begin{figure}[h!]
    \centering
    \includegraphics[width=0.6\textwidth]{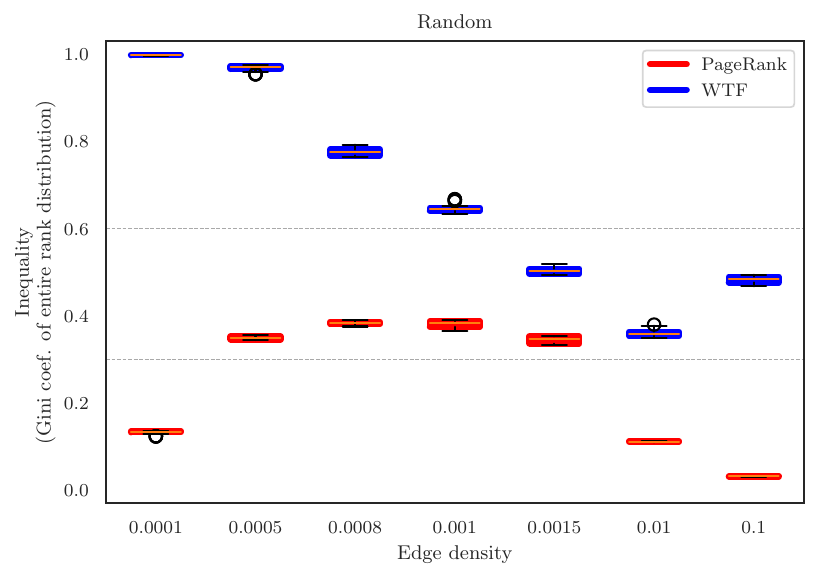}
    \caption{\textbf{Global inequality on Random networks as a function of edge density.} We generate directed Erd\H{o}s-R\'{e}nyi networks to demonstrate how the global inequality (y-axis) varies with respect to the edge density (x-axis) of the network. 
    For each density value we generate networks with different fractions of minorities and $10$ epochs. 
    Note that $d=0.0015$ corresponds to the Random networks used in the main experiments.
    Inequality computed on the PageRank distribution is shown in red, while the inequality on WTF is shown in blue.
    We see different trends for each algorithm. First, the inequality (Gini coefficient) of PageRank is very low when the edge density is extreme (i.e., either too low or too high). This means that in these regimes most nodes are similarly important regardless of the magnitude of their degrees.
    Second, the inequality of WTF is in general negatively correlated with density (i.e., the lower the density, the higher the inequality \cite{goswami2018sparsity}). However, in the extreme case of denser networks (i.e., $d=0.1$), inequality raises.
    Recall that ranking \textit{inequity} is very close to zero ($ME\approx0$) in random networks. 
    Further studies are required to analytically understand the limits of inequality with respect to density.}
    \label{sm:v_density}
\end{figure}

\begin{figure}[ht]
    \centering
    \includegraphics[width=0.6\textwidth]{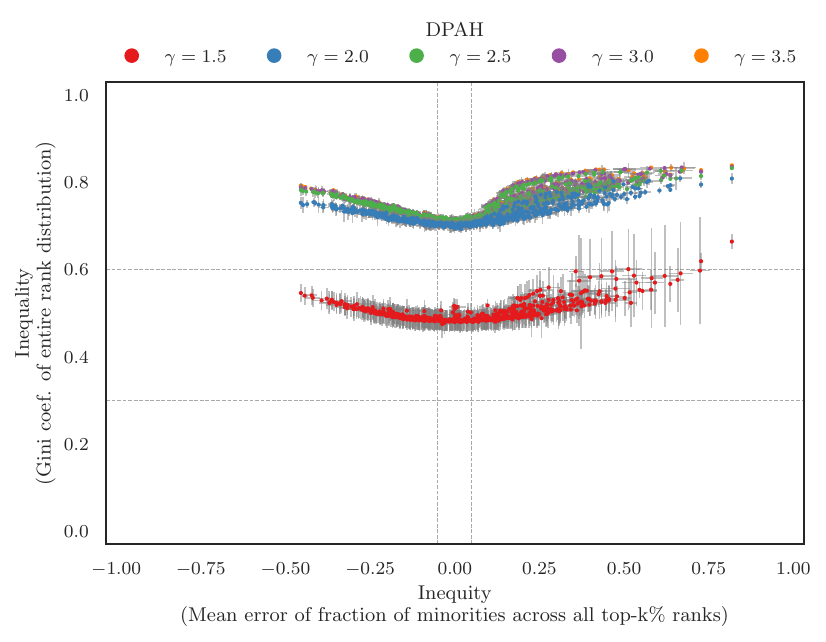}
    \caption{\textbf{The effects of symmetric node activity in the global disparity of PageRank.} 
    We generate {DPAH}~networks by varying $f_m$, $h_{MM}$, $h_{mm}$, and fixing number of nodes $n=2000$ and edge density $d=0.0015$.
    Each color represents the activity of nodes as the out-degree exponents of the networks $\gamma=\gamma_M=\gamma_m\in\{1.5, 2.0, 2.5, 3.0, 3.5\}$.
    We see that by reducing the out-degree exponent in the {DPAH}~networks (from $\gamma=3.5$ to $\gamma=1.5$), we reduce inequality (vertical axis), and inequity remains stable. 
    Vertical and horizontal error bars represent the standard deviation over 10 runs of the Gini and ME, respectively.}
    \label{sm:vh_pr}
\end{figure}

\begin{figure}[ht]
    \centering
    \includegraphics[width=0.6\textwidth]{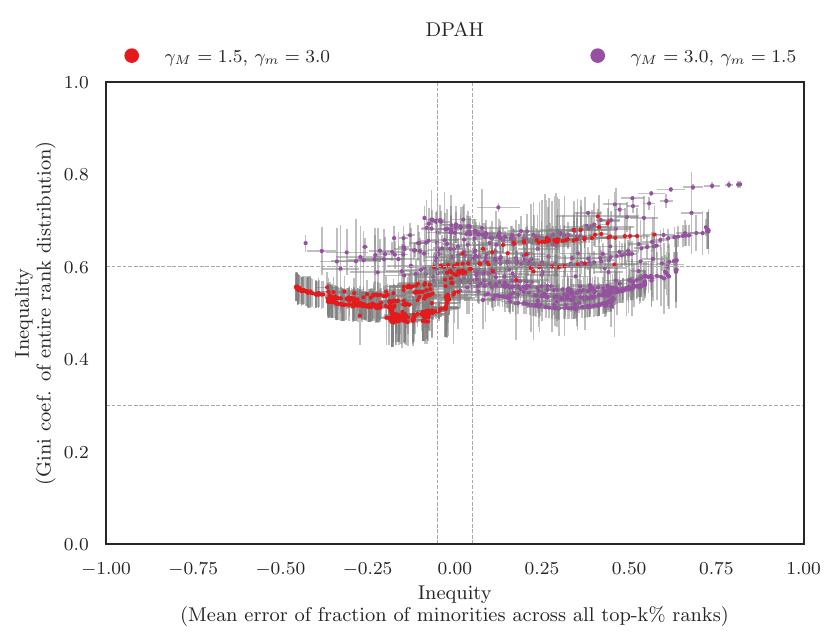}
    \caption{\hl{\textbf{The effects of asymmetric node activity in the global disparity of PageRank.} 
    Similar to} \Cref{sm:vh_pr},\hl{ we generate {DPAH}~networks by varying $f_m$, $h_{MM}$, $h_{mm}$ and fixing number of nodes $n=2000$ and edge density $d=0.0015$. Additionally, we adjust $\gamma_M$ and $\gamma_m$, the activity of majority and minority groups, respectively.
    Red represents networks where the majority is more active (higher out-degree) than the minority. 
    In contrast, purple represents networks where the minority is more active than the majority. 
    In comparison with their counterpart $\gamma=\gamma_M=\gamma_m=3.0$ in} \Cref{sm:vh_pr}, \hl{we see that a more active minority ($\gamma_M=3.0$, $\gamma_m=1.5$), reduces inequality (lower Gini) by amplifying its visibility in the rank (positive ME). Conversely, a more active majority ($\gamma_M=1.5$, $\gamma_m=3.0$) can amplify minority representation at the cost of increasing inequality. However, in general, active majorities benefit themselves in the rank. 
    Vertical and horizontal error bars represent the standard deviation over 10 runs of the Gini and ME, respectively.}}
    \label{sm:vh_pr_agamma}
\end{figure}

\begin{figure}[h!]
    \centering
    \includegraphics[width=\textwidth]{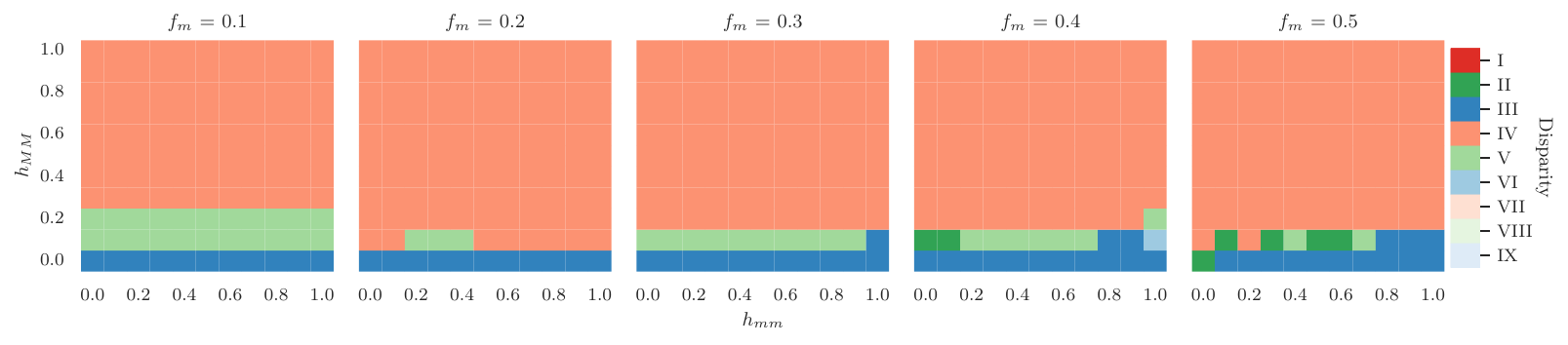}
    \caption{\hl{\textbf{Disparities caused by an active majority group on PageRank as a function of homophily and group size.} Columns represent the fraction of minorities in the network, x-axis indicates the homophily within minorities, and y-axis the homophily within majorities. 
    Colors denote the regions where the disparity lies according to our interpretation (see Figure 2 in main article).
    These results correspond to the red data points in} \Cref{{sm:vh_pr_agamma}}, \hl{$\gamma_M=1.5 < \gamma_m=3.0$. We see that when the majority is more active than the minority, the algorithm reduces the representation of minorities in almost all combinations of homophily and group size. The exception lies in extreme heterophilic majorities. In this case, the ranking may amplify or replicate the visibility of minorities in the rank.}}
    \label{sm:vh_pagerank_DBAH_active_maj}
\end{figure}

\begin{figure}[h!]
    \centering
    \includegraphics[width=\textwidth]{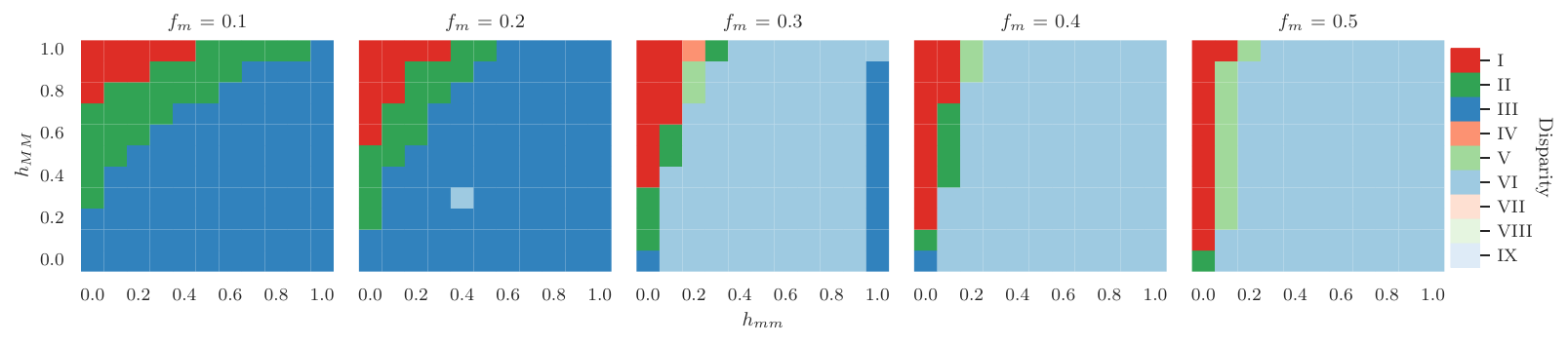}
    \caption{\hl{\textbf{Disparities caused by an active minority group on PageRank as a function of homophily and group size.} Columns represent the fraction of minorities in the network, x-axis indicates the homophily within minorities, and y-axis the homophily within majorities. 
    Colors denote the regions where the disparity lies according to our interpretation (see Figure 2 in main article).
    These results correspond to the purple data points in} \Cref{{sm:vh_pr_agamma}}, \hl{$\gamma_M=3.0 > \gamma_m=1.5$. We see that when the minority is more active than the majority, the algorithm amplifies the representation of minorities in the rank in almost all combinations of homophily and group size. The exception is an interplay between the fraction of minorities and the homophily of the majority group when minorities are heterophilic.}}
    \label{sm:vh_pagerank_DBAH_active_min}
\end{figure}

\begin{figure}[h!]
    \centering
    \includegraphics[width=1\textwidth]{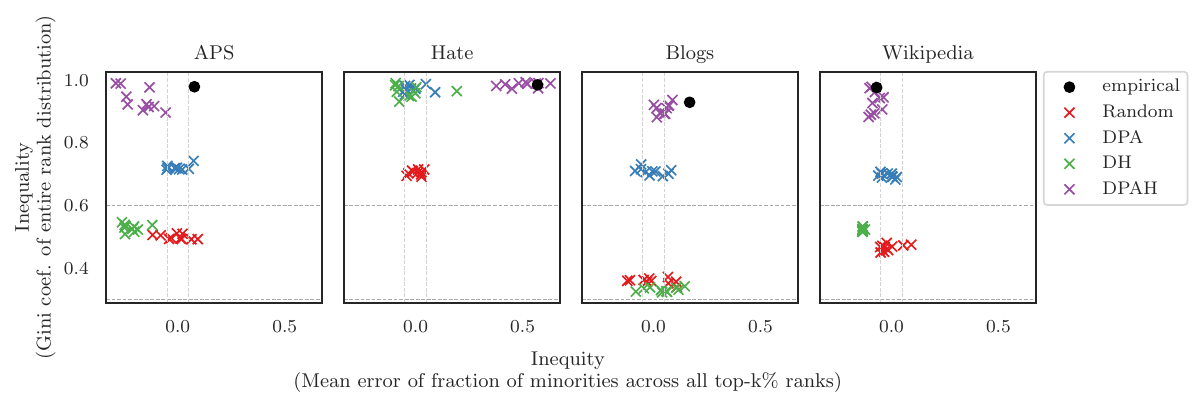}
    \caption{\textbf{Global disparity in WTF on empirical networks.} Each column represents an empirical network. Citation/retweet networks (APS and Hate) and Hyper-link networks (Blogs and Wikipedia). Inequality and inequity are shown in the y- and x-axis, respectively. The disparities in ranking that we see in these empirical networks are best explained by preferential attachment and homophily {DPAH}.}
    \label{sm:empirical}
\end{figure}

\begin{thebibliography}{10}
\urlstyle{rm}
\expandafter\ifx\csname url\endcsname\relax
  \def\url#1{\texttt{#1}}\fi
\expandafter\ifx\csname urlprefix\endcsname\relax\def\urlprefix{URL }\fi
\expandafter\ifx\csname doiprefix\endcsname\relax\def\doiprefix{DOI: }\fi
\providecommand{\bibinfo}[2]{#2}
\providecommand{\eprint}[2][]{\url{#2}}

\bibitem{burt1976positions}
\bibinfo{author}{Burt, R.~S.}
\newblock \bibinfo{journal}{\bibinfo{title}{Positions in networks}}.
\newblock {\emph{\JournalTitle{Social forces}}} \textbf{\bibinfo{volume}{55}},
  \bibinfo{pages}{93--122} (\bibinfo{year}{1976}).

\bibitem{coleman1988social}
\bibinfo{author}{Coleman, J.~S.}
\newblock \bibinfo{journal}{\bibinfo{title}{Social capital in the creation of
  human capital}}.
\newblock {\emph{\JournalTitle{American journal of sociology}}}
  \textbf{\bibinfo{volume}{94}}, \bibinfo{pages}{S95--S120}
  (\bibinfo{year}{1988}).

\bibitem{burt2003social}
\bibinfo{author}{Burt, R.~S.}
\newblock \bibinfo{journal}{\bibinfo{title}{The social structure of
  competition}}.
\newblock {\emph{\JournalTitle{Networks in the knowledge economy}}}
  \textbf{\bibinfo{volume}{13}}, \bibinfo{pages}{57--91}
  (\bibinfo{year}{2003}).

\bibitem{morselli2003career}
\bibinfo{author}{Morselli, C.}
\newblock \bibinfo{journal}{\bibinfo{title}{Career opportunities and
  network-based privileges in the cosa nostra}}.
\newblock {\emph{\JournalTitle{Crime, Law and Social Change}}}
  \textbf{\bibinfo{volume}{39}}, \bibinfo{pages}{383--418}
  (\bibinfo{year}{2003}).

\bibitem{bottero2011worlds}
\bibinfo{author}{Bottero, W.} \& \bibinfo{author}{Crossley, N.}
\newblock \bibinfo{journal}{\bibinfo{title}{Worlds, fields and networks:
  Becker, bourdieu and the structures of social relations}}.
\newblock {\emph{\JournalTitle{Cultural sociology}}}
  \textbf{\bibinfo{volume}{5}}, \bibinfo{pages}{99--119}
  (\bibinfo{year}{2011}).

\bibitem{espin2021explaining}
\bibinfo{author}{Esp{\'\i}n-Noboa, L.}, \bibinfo{author}{Karimi, F.},
  \bibinfo{author}{Ribeiro, B.}, \bibinfo{author}{Lerman, K.} \&
  \bibinfo{author}{Wagner, C.}
\newblock \bibinfo{journal}{\bibinfo{title}{Explaining classification
  performance and bias via network structure and sampling technique}}.
\newblock {\emph{\JournalTitle{Applied Network Science}}}
  \textbf{\bibinfo{volume}{6}}, \bibinfo{pages}{1--25} (\bibinfo{year}{2021}).

\bibitem{abdollahpouri2019unfairness}
\bibinfo{author}{Abdollahpouri, H.}, \bibinfo{author}{Mansoury, M.},
  \bibinfo{author}{Burke, R.} \& \bibinfo{author}{Mobasher, B.}
\newblock \bibinfo{journal}{\bibinfo{title}{The unfairness of popularity bias
  in recommendation}}.
\newblock {\emph{\JournalTitle{arXiv preprint arXiv:1907.13286}}}
  \textbf{\bibinfo{volume}{2440}} (\bibinfo{year}{2019}).
\newblock \bibinfo{note}{[Online; accessed 02-June-2021]}.

\bibitem{gupta2013wtf}
\bibinfo{author}{Gupta, P.} \emph{et~al.}
\newblock \bibinfo{title}{Wtf: The who to follow service at twitter}.
\newblock In \emph{\bibinfo{booktitle}{Proceedings of the 22nd international
  conference on World Wide Web}}, \bibinfo{pages}{505--514}
  (\bibinfo{year}{2013}).

\bibitem{su2016effect}
\bibinfo{author}{Su, J.}, \bibinfo{author}{Sharma, A.} \&
  \bibinfo{author}{Goel, S.}
\newblock \bibinfo{title}{The effect of recommendations on network structure}.
\newblock In \emph{\bibinfo{booktitle}{Proceedings of the 25th international
  conference on World Wide Web}}, \bibinfo{pages}{1157--1167}
  (\bibinfo{year}{2016}).

\bibitem{bellogin2017statistical}
\bibinfo{author}{Bellog{\'\i}n, A.}, \bibinfo{author}{Castells, P.} \&
  \bibinfo{author}{Cantador, I.}
\newblock \bibinfo{journal}{\bibinfo{title}{Statistical biases in information
  retrieval metrics for recommender systems}}.
\newblock {\emph{\JournalTitle{Information Retrieval Journal}}}
  \textbf{\bibinfo{volume}{20}}, \bibinfo{pages}{606--634}
  (\bibinfo{year}{2017}).

\bibitem{page1999pagerank}
\bibinfo{author}{Page, L.}, \bibinfo{author}{Brin, S.},
  \bibinfo{author}{Motwani, R.} \& \bibinfo{author}{Winograd, T.}
\newblock \bibinfo{title}{The pagerank citation ranking: Bringing order to the
  web.}
\newblock \bibinfo{type}{Tech. Rep.}, \bibinfo{institution}{Stanford InfoLab}
  (\bibinfo{year}{1999}).

\bibitem{ghoshal2011ranking}
\bibinfo{author}{Ghoshal, G.} \& \bibinfo{author}{Barab{\'a}si, A.-L.}
\newblock \bibinfo{journal}{\bibinfo{title}{Ranking stability and super-stable
  nodes in complex networks}}.
\newblock {\emph{\JournalTitle{Nature communications}}}
  \textbf{\bibinfo{volume}{2}}, \bibinfo{pages}{394} (\bibinfo{year}{2011}).

\bibitem{karimi2018homophily}
\bibinfo{author}{Karimi, F.}, \bibinfo{author}{G{\'e}nois, M.},
  \bibinfo{author}{Wagner, C.}, \bibinfo{author}{Singer, P.} \&
  \bibinfo{author}{Strohmaier, M.}
\newblock \bibinfo{journal}{\bibinfo{title}{Homophily influences ranking of
  minorities in social networks}}.
\newblock {\emph{\JournalTitle{Scientific reports}}}
  \textbf{\bibinfo{volume}{8}} (\bibinfo{year}{2018}).

\bibitem{fabbri2020effect}
\bibinfo{author}{Fabbri, F.}, \bibinfo{author}{Bonchi, F.},
  \bibinfo{author}{Boratto, L.} \& \bibinfo{author}{Castillo, C.}
\newblock \bibinfo{title}{The effect of homophily on disparate visibility of
  minorities in people recommender systems}.
\newblock In \emph{\bibinfo{booktitle}{Proceedings of the International AAAI
  Conference on Web and Social Media}}, vol.~\bibinfo{volume}{14},
  \bibinfo{pages}{165--175} (\bibinfo{year}{2020}).

\bibitem{cotter2001glass}
\bibinfo{author}{Cotter, D.~A.}, \bibinfo{author}{Hermsen, J.~M.},
  \bibinfo{author}{Ovadia, S.} \& \bibinfo{author}{Vanneman, R.}
\newblock \bibinfo{journal}{\bibinfo{title}{The glass ceiling effect}}.
\newblock {\emph{\JournalTitle{Social Forces}}} \textbf{\bibinfo{volume}{80}},
  \bibinfo{pages}{655--681}, \doiprefix\url{10.1353/sof.2001.0091}
  (\bibinfo{year}{2001}).
\newblock
  \eprint{https://academic.oup.com/sf/article-pdf/80/2/655/6519837/80-2-655.pdf}.

\bibitem{avin2015homophily}
\bibinfo{author}{Avin, C.} \emph{et~al.}
\newblock \bibinfo{title}{Homophily and the glass ceiling effect in social
  networks}.
\newblock In \emph{\bibinfo{booktitle}{Proceedings of the 2015 conference on
  innovations in theoretical computer science}}, \bibinfo{pages}{41--50}
  (\bibinfo{year}{2015}).

\bibitem{stoica2018algorithmic}
\bibinfo{author}{Stoica, A.-A.}, \bibinfo{author}{Riederer, C.} \&
  \bibinfo{author}{Chaintreau, A.}
\newblock \bibinfo{title}{Algorithmic glass ceiling in social networks: The
  effects of social recommendations on network diversity}.
\newblock In \emph{\bibinfo{booktitle}{Proceedings of the 2018 World Wide Web
  Conference}}, \bibinfo{pages}{923--932} (\bibinfo{year}{2018}).

\bibitem{franklin2000invisibility}
\bibinfo{author}{Franklin, A.~J.} \& \bibinfo{author}{Boyd-Franklin, N.}
\newblock \bibinfo{journal}{\bibinfo{title}{Invisibility syndrome: A clinical
  model of the effects of racism on african-american males}}.
\newblock {\emph{\JournalTitle{American Journal of Orthopsychiatry}}}
  \textbf{\bibinfo{volume}{70}}, \bibinfo{pages}{33--41}
  (\bibinfo{year}{2000}).

\bibitem{zehlike2021fairness}
\bibinfo{author}{Zehlike, M.}, \bibinfo{author}{Yang, K.} \&
  \bibinfo{author}{Stoyanovich, J.}
\newblock \bibinfo{journal}{\bibinfo{title}{Fairness in ranking: A survey}}.
\newblock {\emph{\JournalTitle{arXiv preprint arXiv:2103.14000}}}
  (\bibinfo{year}{2021}).

\bibitem{kleinberg2018selection}
\bibinfo{author}{Kleinberg, J.} \& \bibinfo{author}{Raghavan, M.}
\newblock \bibinfo{journal}{\bibinfo{title}{Selection problems in the presence
  of implicit bias}}.
\newblock {\emph{\JournalTitle{arXiv preprint arXiv:1801.03533}}}
  (\bibinfo{year}{2018}).

\bibitem{asudeh2019designing}
\bibinfo{author}{Asudeh, A.}, \bibinfo{author}{Jagadish, H.},
  \bibinfo{author}{Stoyanovich, J.} \& \bibinfo{author}{Das, G.}
\newblock \bibinfo{title}{Designing fair ranking schemes}.
\newblock In \emph{\bibinfo{booktitle}{Proceedings of the 2019 International
  Conference on Management of Data}}, \bibinfo{pages}{1259--1276}
  (\bibinfo{year}{2019}).

\bibitem{yang2017measuring}
\bibinfo{author}{Yang, K.} \& \bibinfo{author}{Stoyanovich, J.}
\newblock \bibinfo{title}{Measuring fairness in ranked outputs}.
\newblock In \emph{\bibinfo{booktitle}{Proceedings of the 29th International
  Conference on Scientific and Statistical Database Management}},
  \bibinfo{pages}{22} (\bibinfo{organization}{ACM}, \bibinfo{year}{2017}).

\bibitem{borgatti2018analyzing}
\bibinfo{author}{Borgatti, S.~P.}, \bibinfo{author}{Everett, M.~G.} \&
  \bibinfo{author}{Johnson, J.~C.}
\newblock \emph{\bibinfo{title}{Analyzing social networks}}
  (\bibinfo{publisher}{Sage}, \bibinfo{year}{2018}).

\bibitem{mcpherson2001birds}
\bibinfo{author}{McPherson, M.}, \bibinfo{author}{Smith-Lovin, L.} \&
  \bibinfo{author}{Cook, J.~M.}
\newblock \bibinfo{journal}{\bibinfo{title}{Birds of a feather: Homophily in
  social networks}}.
\newblock {\emph{\JournalTitle{Annual review of sociology}}}
  \textbf{\bibinfo{volume}{27}}, \bibinfo{pages}{415--444}
  (\bibinfo{year}{2001}).

\bibitem{barabasi1999emergence}
\bibinfo{author}{Barab{\'a}si, A.-L.} \& \bibinfo{author}{Albert, R.}
\newblock \bibinfo{journal}{\bibinfo{title}{Emergence of scaling in random
  networks}}.
\newblock {\emph{\JournalTitle{science}}} \textbf{\bibinfo{volume}{286}},
  \bibinfo{pages}{509--512} (\bibinfo{year}{1999}).

\bibitem{drosou2017diversity}
\bibinfo{author}{Drosou, M.}, \bibinfo{author}{Jagadish, H.},
  \bibinfo{author}{Pitoura, E.} \& \bibinfo{author}{Stoyanovich, J.}
\newblock \bibinfo{journal}{\bibinfo{title}{Diversity in big data: A review}}.
\newblock {\emph{\JournalTitle{Big data}}} \textbf{\bibinfo{volume}{5}},
  \bibinfo{pages}{73--84} (\bibinfo{year}{2017}).

\bibitem{singh2018fairness}
\bibinfo{author}{Singh, A.} \& \bibinfo{author}{Joachims, T.}
\newblock \bibinfo{title}{Fairness of exposure in rankings}.
\newblock In \emph{\bibinfo{booktitle}{Proceedings of the 24th ACM SIGKDD
  International Conference on Knowledge Discovery \& Data Mining}},
  \bibinfo{pages}{2219--2228} (\bibinfo{organization}{ACM},
  \bibinfo{year}{2018}).

\bibitem{dwork2012fairness}
\bibinfo{author}{Dwork, C.}, \bibinfo{author}{Hardt, M.},
  \bibinfo{author}{Pitassi, T.}, \bibinfo{author}{Reingold, O.} \&
  \bibinfo{author}{Zemel, R.}
\newblock \bibinfo{title}{Fairness through awareness}.
\newblock In \emph{\bibinfo{booktitle}{Proceedings of the 3rd innovations in
  theoretical computer science conference}}, \bibinfo{pages}{214--226}
  (\bibinfo{year}{2012}).

\bibitem{kong2021first}
\bibinfo{author}{Kong, H.}, \bibinfo{author}{Martin-Gutierrez, S.} \&
  \bibinfo{author}{Karimi, F.}
\newblock \bibinfo{journal}{\bibinfo{title}{First-mover advantage explains
  gender disparities in physics citations}}.
\newblock {\emph{\JournalTitle{arXiv preprint arXiv:2110.02815}}}
  (\bibinfo{year}{2021}).

\bibitem{rovira2019ranking}
\bibinfo{author}{Rovira, C.}, \bibinfo{author}{Codina, L.},
  \bibinfo{author}{Guerrero-Sol{\'e}, F.} \& \bibinfo{author}{Lopezosa, C.}
\newblock \bibinfo{journal}{\bibinfo{title}{Ranking by relevance and citation
  counts, a comparative study: Google scholar, microsoft academic, wos and
  scopus}}.
\newblock {\emph{\JournalTitle{Future Internet}}}
  \textbf{\bibinfo{volume}{11}}, \bibinfo{pages}{202} (\bibinfo{year}{2019}).

\bibitem{voitalov2019scale}
\bibinfo{author}{Voitalov, I.}, \bibinfo{author}{van~der Hoorn, P.},
  \bibinfo{author}{van~der Hofstad, R.} \& \bibinfo{author}{Krioukov, D.}
\newblock \bibinfo{journal}{\bibinfo{title}{Scale-free networks well done}}.
\newblock {\emph{\JournalTitle{Physical Review Research}}}
  \textbf{\bibinfo{volume}{1}}, \bibinfo{pages}{033034} (\bibinfo{year}{2019}).

\bibitem{pandurangan2002using}
\bibinfo{author}{Pandurangan, G.}, \bibinfo{author}{Raghavan, P.} \&
  \bibinfo{author}{Upfal, E.}
\newblock \bibinfo{title}{Using pagerank to characterize web structure}.
\newblock In \emph{\bibinfo{booktitle}{International computing and
  combinatorics conference}}, \bibinfo{pages}{330--339}
  (\bibinfo{organization}{Springer}, \bibinfo{year}{2002}).

\bibitem{fortunato2007local}
\bibinfo{author}{Fortunato, S.}, \bibinfo{author}{Bogun{\'a}, M.},
  \bibinfo{author}{Flammini, A.} \& \bibinfo{author}{Menczer, F.}
\newblock \bibinfo{journal}{\bibinfo{title}{On local estimations of pagerank: a
  mean field approach}}.
\newblock {\emph{\JournalTitle{Internet Mathematics}}}
  \textbf{\bibinfo{volume}{4}}, \bibinfo{pages}{245--266}
  (\bibinfo{year}{2007}).

\bibitem{albert2002statistical}
\bibinfo{author}{Albert, R.} \& \bibinfo{author}{Barab{\'a}si, A.-L.}
\newblock \bibinfo{journal}{\bibinfo{title}{Statistical mechanics of complex
  networks}}.
\newblock {\emph{\JournalTitle{Reviews of modern physics}}}
  \textbf{\bibinfo{volume}{74}}, \bibinfo{pages}{47} (\bibinfo{year}{2002}).

\bibitem{chawla2002smote}
\bibinfo{author}{Chawla, N.~V.}, \bibinfo{author}{Bowyer, K.~W.},
  \bibinfo{author}{Hall, L.~O.} \& \bibinfo{author}{Kegelmeyer, W.~P.}
\newblock \bibinfo{journal}{\bibinfo{title}{Smote: synthetic minority
  over-sampling technique}}.
\newblock {\emph{\JournalTitle{Journal of artificial intelligence research}}}
  \textbf{\bibinfo{volume}{16}}, \bibinfo{pages}{321--357}
  (\bibinfo{year}{2002}).

\bibitem{peel2018multiscale}
\bibinfo{author}{Peel, L.}, \bibinfo{author}{Delvenne, J.-C.} \&
  \bibinfo{author}{Lambiotte, R.}
\newblock \bibinfo{journal}{\bibinfo{title}{Multiscale mixing patterns in
  networks}}.
\newblock {\emph{\JournalTitle{Proceedings of the National Academy of
  Sciences}}} \textbf{\bibinfo{volume}{115}}, \bibinfo{pages}{4057--4062}
  (\bibinfo{year}{2018}).

\bibitem{smith1987some}
\bibinfo{author}{Smith, M.}
\newblock \bibinfo{journal}{\bibinfo{title}{Some problems with minority
  concepts and a solution}}.
\newblock {\emph{\JournalTitle{Ethnic and Racial Studies}}}
  \textbf{\bibinfo{volume}{10}}, \bibinfo{pages}{341--362}
  (\bibinfo{year}{1987}).

\bibitem{capotorti1979study}
\bibinfo{author}{Capotorti, F.}
\newblock \emph{\bibinfo{title}{Study on the rights of persons belonging to
  ethnic, religious and linguistic minorities}}, vol. \bibinfo{volume}{384}
  (\bibinfo{publisher}{New York: United Nations}, \bibinfo{year}{1979}).

\bibitem{hannum2007concept}
\bibinfo{author}{Hannum, H.}
\newblock \bibinfo{journal}{\bibinfo{title}{The concept and definition of
  minorities}}.
\newblock {\emph{\JournalTitle{Universal Minority Rights}}} \bibinfo{pages}{49}
  (\bibinfo{year}{2007}).

\bibitem{wirth1945problem}
\bibinfo{author}{Wirth, L.}
\newblock \bibinfo{journal}{\bibinfo{title}{The problem of minority groups (pp.
  347--72)}}.
\newblock {\emph{\JournalTitle{Indianapolis, IN: Bobbs-Merrill}}}
  (\bibinfo{year}{1945}).

\bibitem{hurtado2018intersectional}
\bibinfo{author}{Hurtado, A.}
\newblock \bibinfo{journal}{\bibinfo{title}{Intersectional understandings of
  inequality}}.
\newblock {\emph{\JournalTitle{The Oxford handbook of social psychology and
  social justice}}} \bibinfo{pages}{157--172} (\bibinfo{year}{2018}).

\bibitem{erdos1959random}
\bibinfo{author}{{Erd\"{o}s, Paul, and Alfr\'{e}d, R\'{e}nyi}}.
\newblock \bibinfo{journal}{\bibinfo{title}{On random graphs}}.
\newblock {\emph{\JournalTitle{Publicationes Mathematicae}}}
  \textbf{\bibinfo{volume}{6}}, \bibinfo{pages}{290--297}
  (\bibinfo{year}{1959}).

\bibitem{davis1970clustering}
\bibinfo{author}{Davis, J.~A.}
\newblock \bibinfo{journal}{\bibinfo{title}{Clustering and hierarchy in
  interpersonal relations: Testing two graph theoretical models on 742
  sociomatrices}}.
\newblock {\emph{\JournalTitle{American Sociological Review}}}
  \bibinfo{pages}{843--851} (\bibinfo{year}{1970}).

\bibitem{block2015reciprocity}
\bibinfo{author}{Block, P.}
\newblock \bibinfo{journal}{\bibinfo{title}{Reciprocity, transitivity, and the
  mysterious three-cycle}}.
\newblock {\emph{\JournalTitle{Social Networks}}}
  \textbf{\bibinfo{volume}{40}}, \bibinfo{pages}{163--173}
  (\bibinfo{year}{2015}).

\bibitem{dufwenberg2017reciprocity}
\bibinfo{author}{Dufwenberg, M.} \& \bibinfo{author}{Patel, A.}
\newblock \bibinfo{journal}{\bibinfo{title}{Reciprocity networks and the
  participation problem}}.
\newblock {\emph{\JournalTitle{Games and Economic Behavior}}}
  \textbf{\bibinfo{volume}{101}}, \bibinfo{pages}{260--272}
  (\bibinfo{year}{2017}).

\bibitem{altenburger2018monophily}
\bibinfo{author}{Altenburger, K.~M.} \& \bibinfo{author}{Ugander, J.}
\newblock \bibinfo{journal}{\bibinfo{title}{Monophily in social networks
  introduces similarity among friends-of-friends}}.
\newblock {\emph{\JournalTitle{Nature human behaviour}}}
  \textbf{\bibinfo{volume}{2}}, \bibinfo{pages}{284--290}
  (\bibinfo{year}{2018}).

\bibitem{Evtushenko2021paradox}
\bibinfo{author}{Evtushenko, A.} \& \bibinfo{author}{Kleinberg, J.}
\newblock \bibinfo{journal}{\bibinfo{title}{The paradox of second-order
  homophily in networks}}.
\newblock {\emph{\JournalTitle{Scientific Reports}}}
  \textbf{\bibinfo{volume}{11}}, \doiprefix\url{0.1038/s41598-021-92719-6}
  (\bibinfo{year}{2021}).

\bibitem{holland1983stochastic}
\bibinfo{author}{Holland, P.~W.}, \bibinfo{author}{Laskey, K.~B.} \&
  \bibinfo{author}{Leinhardt, S.}
\newblock \bibinfo{journal}{\bibinfo{title}{Stochastic blockmodels: First
  steps}}.
\newblock {\emph{\JournalTitle{Social networks}}} \textbf{\bibinfo{volume}{5}},
  \bibinfo{pages}{109--137} (\bibinfo{year}{1983}).

\bibitem{newman2003structure}
\bibinfo{author}{Newman, M.~E.}
\newblock \bibinfo{journal}{\bibinfo{title}{The structure and function of
  complex networks}}.
\newblock {\emph{\JournalTitle{SIAM review}}} \textbf{\bibinfo{volume}{45}},
  \bibinfo{pages}{167--256} (\bibinfo{year}{2003}).

\bibitem{merton1988matthew}
\bibinfo{author}{Merton, R.~K.}
\newblock \bibinfo{journal}{\bibinfo{title}{The matthew effect in science, ii:
  Cumulative advantage and the symbolism of intellectual property}}.
\newblock {\emph{\JournalTitle{isis}}} \textbf{\bibinfo{volume}{79}},
  \bibinfo{pages}{606--623} (\bibinfo{year}{1988}).

\bibitem{perra2012activity}
\bibinfo{author}{Perra, N.}, \bibinfo{author}{Gon{\c{c}}alves, B.},
  \bibinfo{author}{Pastor-Satorras, R.} \& \bibinfo{author}{Vespignani, A.}
\newblock \bibinfo{journal}{\bibinfo{title}{Activity driven modeling of time
  varying networks}}.
\newblock {\emph{\JournalTitle{Scientific reports}}}
  \textbf{\bibinfo{volume}{2}}, \bibinfo{pages}{469} (\bibinfo{year}{2012}).

\bibitem{rogers1970homophily}
\bibinfo{author}{Rogers, E.~M.} \& \bibinfo{author}{Bhowmik, D.~K.}
\newblock \bibinfo{journal}{\bibinfo{title}{Homophily-heterophily: Relational
  concepts for communication research}}.
\newblock {\emph{\JournalTitle{Public opinion quarterly}}}
  \textbf{\bibinfo{volume}{34}}, \bibinfo{pages}{523--538}
  (\bibinfo{year}{1970}).

\bibitem{kaggle:hate}
\bibinfo{author}{Ribeiro, M.}
\newblock \bibinfo{title}{Hateful users on twitter: Detecting hate speech with
  context.}
\newblock
  \bibinfo{howpublished}{\url{https://www.kaggle.com/manoelribeiro/hateful-users-on-twitter}}
  (\bibinfo{year}{2018}).
\newblock \bibinfo{note}{Accessed: 2020-10-07}.

\bibitem{adamic2005political}
\bibinfo{author}{Adamic, L.~A.} \& \bibinfo{author}{Glance, N.}
\newblock \bibinfo{title}{The political blogosphere and the 2004 us election:
  divided they blog}.
\newblock In \emph{\bibinfo{booktitle}{Proceedings of the 3rd international
  workshop on Link discovery}}, \bibinfo{pages}{36--43}
  (\bibinfo{organization}{ACM}, \bibinfo{year}{2005}).

\bibitem{wikiwag}
\bibinfo{author}{Wagner, C.}
\newblock \bibinfo{title}{Politicians on wikipedia and dbpedia (version:
  1.0.0)}, \doiprefix\url{https://doi.org/10.7802/1515} (\bibinfo{year}{2017}).

\bibitem{wiki:poli}
\bibinfo{author}{GESIS}.
\newblock \bibinfo{title}{Temporal network of politicians on wikipedia}.
\newblock
  \bibinfo{howpublished}{\url{https://github.com/gesiscss/Wikipedia-Politician-Network}}
  (\bibinfo{year}{2018}).
\newblock \bibinfo{note}{Accessed: 2020-10-07}.

\bibitem{lofgren2016personalized}
\bibinfo{author}{Lofgren, P.}, \bibinfo{author}{Banerjee, S.} \&
  \bibinfo{author}{Goel, A.}
\newblock \bibinfo{title}{Personalized pagerank estimation and search: A
  bidirectional approach}.
\newblock In \emph{\bibinfo{booktitle}{Proceedings of the Ninth ACM
  International Conference on Web Search and Data Mining}},
  \bibinfo{pages}{163--172} (\bibinfo{year}{2016}).

\bibitem{ding2009pagerank}
\bibinfo{author}{Ding, Y.}, \bibinfo{author}{Yan, E.}, \bibinfo{author}{Frazho,
  A.} \& \bibinfo{author}{Caverlee, J.}
\newblock \bibinfo{journal}{\bibinfo{title}{Pagerank for ranking authors in
  co-citation networks}}.
\newblock {\emph{\JournalTitle{Journal of the American Society for Information
  Science and Technology}}} \textbf{\bibinfo{volume}{60}},
  \bibinfo{pages}{2229--2243} (\bibinfo{year}{2009}).

\bibitem{gollapalli2011ranking}
\bibinfo{author}{Gollapalli, S.~D.}, \bibinfo{author}{Mitra, P.} \&
  \bibinfo{author}{Giles, C.~L.}
\newblock \bibinfo{title}{Ranking authors in digital libraries}.
\newblock In \emph{\bibinfo{booktitle}{Proceedings of the 11th annual
  international ACM/IEEE joint conference on Digital libraries}},
  \bibinfo{pages}{251--254} (\bibinfo{year}{2011}).

\bibitem{senanayake2015pagerank}
\bibinfo{author}{Senanayake, U.}, \bibinfo{author}{Piraveenan, M.} \&
  \bibinfo{author}{Zomaya, A.}
\newblock \bibinfo{journal}{\bibinfo{title}{The pagerank-index: Going beyond
  citation counts in quantifying scientific impact of researchers}}.
\newblock {\emph{\JournalTitle{PloS one}}} \textbf{\bibinfo{volume}{10}},
  \bibinfo{pages}{e0134794} (\bibinfo{year}{2015}).

\bibitem{barbieri2014follow}
\bibinfo{author}{Barbieri, N.}, \bibinfo{author}{Bonchi, F.} \&
  \bibinfo{author}{Manco, G.}
\newblock \bibinfo{title}{Who to follow and why: link prediction with
  explanations}.
\newblock In \emph{\bibinfo{booktitle}{Proceedings of the 20th ACM SIGKDD
  international conference on Knowledge discovery and data mining}},
  \bibinfo{pages}{1266--1275} (\bibinfo{year}{2014}).

\bibitem{yu2014link}
\bibinfo{author}{Yu, Y.} \& \bibinfo{author}{Wang, X.}
\newblock \bibinfo{journal}{\bibinfo{title}{Link prediction in directed network
  and its application in microblog}}.
\newblock {\emph{\JournalTitle{Mathematical Problems in Engineering}}}
  \textbf{\bibinfo{volume}{2014}} (\bibinfo{year}{2014}).

\bibitem{morone2015influence}
\bibinfo{author}{Morone, F.} \& \bibinfo{author}{Makse, H.~A.}
\newblock \bibinfo{journal}{\bibinfo{title}{Influence maximization in complex
  networks through optimal percolation}}.
\newblock {\emph{\JournalTitle{Nature}}} \textbf{\bibinfo{volume}{524}},
  \bibinfo{pages}{65--68} (\bibinfo{year}{2015}).

\bibitem{zhang2016identifying}
\bibinfo{author}{Zhang, J.-X.}, \bibinfo{author}{Chen, D.-B.},
  \bibinfo{author}{Dong, Q.} \& \bibinfo{author}{Zhao, Z.-D.}
\newblock \bibinfo{journal}{\bibinfo{title}{Identifying a set of influential
  spreaders in complex networks}}.
\newblock {\emph{\JournalTitle{Scientific reports}}}
  \textbf{\bibinfo{volume}{6}}, \bibinfo{pages}{27823} (\bibinfo{year}{2016}).

\bibitem{gleich2015pagerank}
\bibinfo{author}{Gleich, D.~F.}
\newblock \bibinfo{journal}{\bibinfo{title}{Pagerank beyond the web}}.
\newblock {\emph{\JournalTitle{SIAM Review}}} \textbf{\bibinfo{volume}{57}},
  \bibinfo{pages}{321--363} (\bibinfo{year}{2015}).

\bibitem{liu2005co}
\bibinfo{author}{Liu, X.}, \bibinfo{author}{Bollen, J.},
  \bibinfo{author}{Nelson, M.~L.} \& \bibinfo{author}{Van~de Sompel, H.}
\newblock \bibinfo{journal}{\bibinfo{title}{Co-authorship networks in the
  digital library research community}}.
\newblock {\emph{\JournalTitle{Information processing \& management}}}
  \textbf{\bibinfo{volume}{41}}, \bibinfo{pages}{1462--1480}
  (\bibinfo{year}{2005}).

\bibitem{jezek2008exploration}
\bibinfo{author}{Jezek, K.}, \bibinfo{author}{Fiala, D.} \&
  \bibinfo{author}{Steinberger, J.}
\newblock \bibinfo{title}{Exploration and evaluation of citation networks.}
\newblock In \emph{\bibinfo{booktitle}{ELPUB}}, \bibinfo{pages}{351--362}
  (\bibinfo{year}{2008}).

\bibitem{fiala2008pagerank}
\bibinfo{author}{Fiala, D.}, \bibinfo{author}{Rousselot, F.} \&
  \bibinfo{author}{Je{\v{z}}ek, K.}
\newblock \bibinfo{journal}{\bibinfo{title}{Pagerank for bibliographic
  networks}}.
\newblock {\emph{\JournalTitle{Scientometrics}}} \textbf{\bibinfo{volume}{76}},
  \bibinfo{pages}{135--158} (\bibinfo{year}{2008}).

\bibitem{brin1998anatomy}
\bibinfo{author}{Brin, S.} \& \bibinfo{author}{Page, L.}
\newblock \bibinfo{journal}{\bibinfo{title}{The anatomy of a large-scale
  hypertextual web search engine}}.
\newblock {\emph{\JournalTitle{Computer networks and ISDN systems}}}
  \textbf{\bibinfo{volume}{30}}, \bibinfo{pages}{107--117}
  (\bibinfo{year}{1998}).

\bibitem{fast-pagerank}
\bibinfo{author}{Sajadi, A.}
\newblock \bibinfo{title}{Fast personalized pagerank implementation.}
\newblock
  \bibinfo{howpublished}{\url{https://github.com/asajadi/fast-pagerank}}
  (\bibinfo{year}{2019}).
\newblock \bibinfo{note}{Accessed: 2021-03-30}.

\bibitem{lempel2001salsa}
\bibinfo{author}{Lempel, R.} \& \bibinfo{author}{Moran, S.}
\newblock \bibinfo{journal}{\bibinfo{title}{Salsa: the stochastic approach for
  link-structure analysis}}.
\newblock {\emph{\JournalTitle{ACM Transactions on Information Systems
  (TOIS)}}} \textbf{\bibinfo{volume}{19}}, \bibinfo{pages}{131--160}
  (\bibinfo{year}{2001}).

\bibitem{jeh2003scaling}
\bibinfo{author}{Jeh, G.} \& \bibinfo{author}{Widom, J.}
\newblock \bibinfo{title}{Scaling personalized web search}.
\newblock In \emph{\bibinfo{booktitle}{Proceedings of the 12th international
  conference on World Wide Web}}, \bibinfo{pages}{271--279}
  (\bibinfo{year}{2003}).

\bibitem{gini1912variabilita}
\bibinfo{author}{Gini, C.}
\newblock \bibinfo{journal}{\bibinfo{title}{Variabilit{\`a} e mutabilit{\`a}}}.
\newblock {\emph{\JournalTitle{Reprinted in Memorie di metodologica statistica
  (Ed. Pizetti E}}}  (\bibinfo{year}{1912}).

\bibitem{ceriani2012origins}
\bibinfo{author}{Ceriani, L.} \& \bibinfo{author}{Verme, P.}
\newblock \bibinfo{journal}{\bibinfo{title}{The origins of the gini index:
  extracts from variabilit{\`a} e mutabilit{\`a} (1912) by corrado gini}}.
\newblock {\emph{\JournalTitle{The Journal of Economic Inequality}}}
  \textbf{\bibinfo{volume}{10}}, \bibinfo{pages}{421--443}
  (\bibinfo{year}{2012}).

\bibitem{gini}
\bibinfo{author}{StatsDirect}.
\newblock \bibinfo{title}{Gini coefficient of inequality}.
\newblock
  \bibinfo{howpublished}{\url{https://www.statsdirect.com/help/default.htm\#nonparametric\_methods/gini.htm}}.
\newblock \bibinfo{note}{Accessed: 2020-11-09}.

\bibitem{scikit-learn}
\bibinfo{author}{Pedregosa, F.} \emph{et~al.}
\newblock \bibinfo{journal}{\bibinfo{title}{Scikit-learn: Machine learning in
  {P}ython}}.
\newblock {\emph{\JournalTitle{Journal of Machine Learning Research}}}
  \textbf{\bibinfo{volume}{12}}, \bibinfo{pages}{2825--2830}
  (\bibinfo{year}{2011}).

\bibitem{randomforest}
\bibinfo{author}{Scikit-learn}.
\newblock \bibinfo{title}{sklearn.ensemble.randomforestregressor}.
\newblock
  \bibinfo{howpublished}{\url{https://scikit-learn.org/stable/modules/generated/sklearn.ensemble.RandomForestRegressor.html}}.
\newblock \bibinfo{note}{Accessed: 2020-11-09}.

\bibitem{fortunato2006approximating}
\bibinfo{author}{Fortunato, S.}, \bibinfo{author}{Bogu{\~n}{\'a}, M.},
  \bibinfo{author}{Flammini, A.} \& \bibinfo{author}{Menczer, F.}
\newblock \bibinfo{title}{Approximating pagerank from in-degree}.
\newblock In \emph{\bibinfo{booktitle}{International workshop on algorithms and
  models for the web-graph}}, \bibinfo{pages}{59--71}
  (\bibinfo{organization}{Springer}, \bibinfo{year}{2006}).

\bibitem{mariani2015ranking}
\bibinfo{author}{Mariani, M.~S.}, \bibinfo{author}{Medo, M.} \&
  \bibinfo{author}{Zhang, Y.-C.}
\newblock \bibinfo{journal}{\bibinfo{title}{Ranking nodes in growing networks:
  When pagerank fails}}.
\newblock {\emph{\JournalTitle{Scientific reports}}}
  \textbf{\bibinfo{volume}{5}}, \bibinfo{pages}{1--10} (\bibinfo{year}{2015}).

\bibitem{goswami2018sparsity}
\bibinfo{author}{Goswami, S.}, \bibinfo{author}{Murthy, C.} \&
  \bibinfo{author}{Das, A.~K.}
\newblock \bibinfo{journal}{\bibinfo{title}{Sparsity measure of a network
  graph: Gini index}}.
\newblock {\emph{\JournalTitle{Information Sciences}}}
  \textbf{\bibinfo{volume}{462}}, \bibinfo{pages}{16--39}
  (\bibinfo{year}{2018}).

\end{thebibliography}
\end{document}